\documentclass{emulateapj}

\usepackage{rotating}

\newcommand{\arcs}{$^{\prime\prime}$}
\newcommand{\Mo}{$~\mathrm{M_\odot}$}

\shorttitle{A catalog of visual double and multiple stars with eclipsing components}
\shortauthors{Zasche et al.}

\begin{document}

\title{A catalog of visual double and multiple stars with eclipsing components}

\author{P. Zasche\altaffilmark{1},\altaffilmark{2} and M. Wolf\altaffilmark{1}}
\affil{\altaffilmark{1} Astronomical Institute, Faculty of Mathematics and Physics, Charles
University of Prague, CZ-180 00 Praha 8, V Hole\v{s}ovi\v{c}k\'ach~2, Czech Republic}
\affil{\altaffilmark{2} Instituto de Astronom\'{\i}a, Universidad Nacional Aut\'onoma de M\'exico,
A.P. 70-264, M\'exico, DF 04510, Mexico}

\author{W.I. Hartkopf\altaffilmark{3}}
\affil{\altaffilmark{3} U.S. Naval Observatory, 3450 Massachusetts Avenue, Washington, DC 20392,
USA}

\author{P. Svoboda\altaffilmark{4}}
\affil{\altaffilmark{4} Private observatory, CZ-614 00 Brno, V\'ypustky 5, Czech Republic}

\author{R. Uhla\v{r}\altaffilmark{5}}
\affil{\altaffilmark{5} Private observatory, Poho\v{r}\'{\i} 71, CZ-25401, J\'{\i}lov\'e u Prahy,
Czech Republic}

\and

\author{A.Liakos\altaffilmark{6} and K. Gazeas\altaffilmark{6}}
\affil{\altaffilmark{6} Department of Astrophysics, Astronomy and Mechanics, Faculty of Physics,
University of Athens, GR-15784 Zografos, Athens, Greece}

\begin{abstract}
\noindent A new catalog of visual double systems containing eclipsing binaries as one component is
presented. The main purpose of this catalog is to compile a complete list of all known multiples of
this variety, both for current analysis and to highlight those in need of additional observations.
All available photometric and astrometric data were analyzed, resulting in new orbits for eight
systems and new times of minimum light for a number of the eclipsing binaries. Some of the systems
in the catalog have acceptable solutions for their visual orbits, although in most cases their
orbital periods are too long for simultaneous analysis. Also included, however, are a number of
systems which currently lack an orbital solution but which may be suitable for simultaneous
analysis in the future.
\end{abstract}


\section{Introduction}

Binary stars are essential objects for determining precise physical properties of stars, especially
masses, through a combined analysis of photometric, astrometric, and spectroscopic data. If the
stars comprise an eclipsing binary (hereafter EB), radii, distance and, in favorable cases,
effective temperatures may also be determined from a combined analysis of light and radial velocity
(hereafter RV) curves. Moreover, masses and distances for visual binaries may be determined from a
combined analysis of astrometric and radial velocity measurements. Additional components may often
be revealed through these analyses; one especially productive source is the study of the long-time
behavior of the period of an EB. As might be expected, the longer the time span of conjunction time
measurements, or times of minimum light, the greater the chance of detecting a long period orbit
due to an additional member of the system under study.

There are currently more than 2000 systems with known visual orbits (see the USNO Sixth Orbit
Catalog\footnote{\url{http://ad.usno.navy.mil/wds/orb6.html}}). Among these systems, at least 34
have been found to include an EB as one of their components. About 100 another EBs were found to be
members of visual pairs or multiples, the orbit of which has not been computed yet. Collecting and
investigating this fraction of the visual doubles is the purpose of this catalog.

The systems presented here were found through searches in the ``Washington Double Star
Catalog"\footnote{http://ad.usno.navy.mil/wds/} (WDS, \citealt{WDS}), identifying systems with EBs
amongst their components. This catalog differs slightly from the one published by
\cite{Chambliss1992}, where all of the multiple-star systems containing EBs known by the author
were collected. \citeauthor{Chambliss1992} mentioned that 80 EBs were known to be components of
multiple-star systems and presented 37 of them in detail. The present catalog is restricted to
reasonably well-observed visual binaries which contain EBs as a component (the qualifier
``reasonably well-observed" being described below). The number of known EBs amongst visual pairs is
growing rapidly; we felt this justified collecting them into a separate catalog.

The $O-C$ diagrams constructed against the EBs linear ephemeris frequently display variations in
their orbital periods (see, for example, a catalog of $O-C$ diagrams of such systems by
\citealt{Kreiner2001}). For a discussion of the details and limitations of $O-C$ diagram analysis,
see \cite{Sterken2005}. A periodic oscillation in an $O-C$ diagram may be explained as a light-time
effect (hereafter LITE) caused by a distant companion orbiting around a common center of mass with
the EB; see \cite{Irwin1959}, or \cite{Mayer1990} for details. In favorable cases, this component
may be identified as a distant member of the system, directly detectable astrometrically as a
visual or interferometric companion. Despite the large number of eclipsing binaries in our Galaxy
(estimated at about 10$^8$, according to \citealt{Kopal1978} and \citealt{Cooper1994}), there are
still only a few dozen systems known where the EBs are members of spatially resolvable pairs. A few
of the most well-known systems are V505~Sgr (see \citealt{Mayer1997}), V819~Her
\citep{Muterspaugh2006}, and VW~Cep (\citealt{ZascheWolf}, hereafter ZW, and \citealt{Zasche2008}).
Decades' worth of data are available for all these systems, making possible an analysis of period
variations and also a solution of the visual orbits (fortunately the visual orbital periods in
these examples are rather short). Despite this fact, until now the results from different
techniques have been in contradiction with each other. A typical example of such a discrepancy is
the system V505~Sgr, where \cite{V505Sgr2006} presented the period of the third body about 44.6 or
38.6~yr from period analysis, while \cite{Cvetkovic2008} derived the third-body period about
60.1~yr from the visual orbit. To the best of our knowledge, the system VW Cep is presently the
most suitable for a combined analysis of variation of minima timings; the combination of available
photometric and interferometric data can even be used to determine the distance to the system. The
distance derived by this method is more precise than any previously derived value; see ZW for
details. Such a combined approach is potentially very powerful, especially anticipating the
high-quality data expected to come from planned astrometric and photometric space missions. The
most serious weakness in this combined method seems to be due to the period of the distant
component, which in most cases is extremely long (typically decades to centuries or longer). As a
result, the data frequently cover only a small arc of the orbit, and this very incomplete coverage
obviously degrades the precision of the results. For a detailed description of the method,
including algorithms, limitations, and results, see ZW and \citealt{Zasche2008}.

{\small {
\begin{sidewaystable*}
\centering \scriptsize
\begin{tabular}{ r l | c c c r r c r c r r r l l c l }
 \hline
 \multicolumn{2}{c|}{Star} & \multicolumn{3}{c}{Spectral types} & \colhead{$V$} &
 \colhead{$P$} & \colhead{$p_3$} & \colhead{$\pi$} & \colhead{Comp} & \colhead{Min} & \colhead{Min}
 & \colhead{$M$} & \colhead{Depth} & \colhead{Depth} & \colhead{Orbit} & \colhead{References} \\
  \colhead{HD} & Designation & 1 & 2 & 3 & \colhead{[mag]} & \colhead{[day]} & \colhead{[yr]} & \colhead{[mas]} &
 \colhead{ }& \colhead{Pri} & \colhead{Sec} & \colhead{Astr.} & \colhead{MinP} & \colhead{MinS} & &  \\
 \multicolumn{1}{c}{(1)} & \multicolumn{1}{c}{(2)} & (3) & (4) & (5) & (6)\phn & \phn(7)\phn\phn\phn
 & (8) & \multicolumn{1}{c}{(9)}\phn & (10) & (11) & (12) & (13) & \multicolumn{1}{c}{(14)} &
 \multicolumn{1}{c}{(15)} & (16) & \multicolumn{1}{c}{(17)} \\ \hline
   123  &  V640~Cas   &   G3V   &     M2-3    &   G9V   &  5.93    &  1.026\phn\phn & \phn106.7\phn\phn           & 49.30 $\pm$ 1.05 & 3 &            2 &  0 & 572   & 0.066 V  &         & y & 1,2,3,4,5 \\
  1082  &  V348~And   &   \multicolumn{3}{c}{ B9IV }    &  6.76    &  5.539\phn\phn & 137.9/330                   &  4.05 $\pm$ 0.76 & 3 &            1 &  0 &  61   & 0.150 Hp &         & y & 1,2,6,7,8 \\
  4134  &  V355~And   &   \multicolumn{3}{c}{ F6V }     &  7.69    &  4.7184\phn    &  ---                        &  8.22 $\pm$ 1.74 & 3 &            7 &  2 &  51   & 0.310 V  & 0.21 V  & n & 1,2,9 \\
  6882  & $\zeta$~Phe &  B6V   &     B8V    &    A7V    &  3.97    &  1.66978       & \phn220.9\phn\phn           & 11.66 $\pm$ 0.77 & 3 &           25 & 17 &  11   & 0.510 V  & 0.31 V  & y & 1,2,11,12,13 \\
  9770  &  BB~Scl     &  K3-4V &   K4-5V    & K1-2V+M2V &  7.14    &  0.47653       &  4.56+111.8                 & 42.29 $\pm$ 1.47 & 4 &            2 &  0 & 88+44 & 0.220 Hp & 0.22 Hp & y & 1,2,14,15,16,17 \\
 10543  &  V773~Cas   & \multicolumn{2}{c}{A3V}&  F0-5  &  6.21    &  1.29367       & \phn304.04\phn              & 12.63 $\pm$ 0.77 & 3 &            4 &  0 &  79   & 0.090 Hp &         & y & 1,2,18,19,20 \\
 12180  &  AA~Cet     &  F2V   &    F2V    &    F5      &  7.22    &  0.53617       &  ---                        &  4.63 $\pm$ 2.36 & 4 &          160 & 80 &  33   & 0.500 p  & 0.50 p  & n & 1,2,21,76 \\
 14817  &  V559~Cas   &  B9V   &    B9V    &    B9V     &  7.02    &  1.5806\phn    & \phn836\phm{.}\phn\phn\phn  &  4.43 $\pm$ 1.51 & 3 &            9 &  6 & 101   & 0.220 V  & 0.20 V  & y & 1,2,22,21 \\
 19356  & $\beta$~Per &  B8V   &    G8IV   &    A7m     &  2.12    &  2.86731       & \phn\phn\phn1.862           & 35.14 $\pm$ 0.90 & 3 &$\approx$1400 &  9 &  37   & 1.270 V  & 0.05 V  & y & 1,2,23,24 \\
 25833  &  AG~Per     &  B3Vn  &     B3    &    B       &  6.69    &  2.02873       &  ---                        &  3.89 $\pm$ 1.31 & 3 &           51 & 52 &  39   & 0.310 V  & 0.31 V  & n & 1,2,25 \\
 29911  &  V592~Per   & \multicolumn{2}{c}{F2IV}&  G0V  &  8.37    &  0.71572       & \phn115.32\phn              &  5.12 $\pm$ 1.55 & 3 &           11 & 15 &  19   & 0.350 Hp & 0.27 Hp & Y & 1,2,26,27 \\
 35411  & $\eta$~Ori  & B1V    &    B3V    &    B2V     &  3.38    &  7.9904\phn    & \phn\phn\phn9.44\phn        &  3.62 $\pm$ 0.88 & 4 &            1 &  0 &  19   & 0.290 V  & 0.26 V  & y & 1,2,21,28 \\
 36486  &$\delta$~Ori~A&09.5II &  B0.5III  &    B       &  2.23    &  5.73248       & \phn704.8\phn\phn           &  3.56 $\pm$ 0.83 & 3 &            6 &  3 &  38   & 0.120 V  & 0.06 V  & Y & 1,2,29 \\
 38735  &  V1031~Ori  & A8III-IV  &  A5IV-V & A6IV-V    &  6.06    &  3.40556       & \phn\phn92.66\phn           &  4.99 $\pm$ 1.04 & 3 &            8 &  5 &  20   & 0.410 V  & 0.30 V  & Y & 1,2,21,30 \\
 57061  &  $\tau$~CMa & \multicolumn{2}{c}{O9II}&  O    &  4.39    &  1.28212       &  ---                        &  1.02 $\pm$ 0.71 & 4 &            2 &  0 &  32   & 0.050 Hp & 0.04 Hp & n & 1,2,31 \\
 60179  &  YY~Gem     &  dM1e  &   dM1e   &     A1V     &  9.07    &  0.81428       &  ---                        & 63.27 $\pm$ 1.23 & 6 &           97 & 85 & 140   & 0.690 V  & 0.68 V  & n & 1,2,32 \\
 66094  &  V635~Mon   & G9III-IV & A2.5   &     F5      &  7.31    &  1.80781       & \phn257\phm{.}\phn\phn\phn  &  3.06 $\pm$ 1.04 & 3 &          113 &  2 &  24   & 0.500 p  &         & y & 1,2,33,34,35 \\
 71487  &  NO~Pup     & B9V    &    A7V   &  A3V+A3V    &  6.7\phn &  1.25688       &  ---                        &  5.32 $\pm$ 0.87 & 5 &           10 &  5 &  29   & 0.450 V  & 0.13 V  & n & 1,2,21,36 \\
 71581  &  VV~Pyx     & \multicolumn{2}{c}{A1V}& A5-A7  &  6.58    &  4.59618       &  ---                        &  4.51 $\pm$ 1.00 & 3 &            6 &  5 &  11   & 0.520 V  & 0.50 V  & n & 1,2,37 \\
 71663  &  LO~Hya     & F0V    &    G0V   &   A5V+F5V   &  6.42    &  2.49963       & \phn\phn54.70\phn           & 11.73 $\pm$ 0.94 & 5 &            3 &  3 &  69   & 0.240 V  & 0.10 V  & y & 1,2,21,38 \\
 74956  & $\delta$~Vel& A0V    &    A1V   &     F       &  1.95    & 45.150\phn\phn & \phn142.00\phn              & 40.90 $\pm$ 0.38 & 3 &            5 &  5 &  38   & 0.400 V  &         & y & 1,2,39,40 \\
 ---~~  &  AC~UMa     & \multicolumn{2}{c}{A2}& G0      & 10.3\phn &  6.85469       & 1199.2\phn\phn              &                  & 3 &           80 &  0 &  10   & 3.700 V  &         & Y & 1,41 \\
 82780  &  DI~Lyn     & F2V    &    F3V   &     G3V     &  6.76    &  1.68154       & \phn\phn64.257              & 11.80 $\pm$ 2.50 & 5 &            3 &  2 &  12   & 0.080 V  & 0.05 V  & y & 1,2,42,43,44 \\
 91636  &  TX~Leo     &  \multicolumn{2}{c}{A2V}&  B    &  5.67    &  2.44507       &  ---                        &  7.05 $\pm$ 0.99 & 3 &            8 &  0 & 140   & 0.090 V  & 0.03 V  & n & 1,2,45,46 \\
 101205 &  V871~Cen   &  \multicolumn{3}{c}{ O7IIIn }   &  6.49    &  2.0842\phn    & 1715.8\phn\phn              & -1.44 $\pm$ 1.42 & 5 &            1 &  0 &  17   & 0.080 V  & 0.08 V  & Y & 1,2,47 \\
 101379 &  GT~Mus     & A0V    &    A2V   & K4III+dF/G  &  5.17    &  2.7546\phn    & \phn\phn90.7\phn\phn        &  5.81 $\pm$ 0.64 & 4 &            0 &  0 &  13   & 0.130 V  &         & y & 1,2,48,49 \\
 103483 &  DN~UMa     & A3V    &    A3V   &     A8-9    &  6.54    &  1.73043       & \phn136.538                 &  4.07 $\pm$ 1.24 & 5 &           11 &  8 &  34   & 0.100 B  & 0.10 B  & y & 1,2,50,51 \\
 110317 &  VV~Crv     &  \multicolumn{3}{c}{ F5IV }     &  5.27    &  3.145\phn\phn &  ---                        & 11.72 $\pm$ 1.90 & 5 &            1 &  0 & 156   & 0.150 Hp &         & n & 1,2,52,53 \\
 114529 &  V831~Cen   & \multicolumn{2}{c}{B8V}&  B9V   &  4.58    &  0.64252       & \phn\phn27.0\phn\phn        &  9.42 $\pm$ 1.52 & 5 &            1 &  0 &  40   & 0.170 V  & 0.15 V  & y & 1,2,54,55 \\
 119931 &  HT~Vir     & F8V    &    F8V   &     F       &  7.16    &  0.40764       & \phn260.7\phn\phn           & 15.39 $\pm$ 2.72 & 4 &           21 & 22 & 277   & 0.420 V  & 0.42 V  & y & 1,2,13,56,57 \\
  ---~~ &  ET~Boo     &  \multicolumn{3}{c}{ F8 }       &  9.09    &  0.64504       & \phn113.32\phn              &  5.85 $\pm$ 1.40 & 4 &           18 & 13 &  20   & 0.300 Hp & 0.20 Hp & y & 1,2,58,59 \\
 133640 &  i~Boo      & K0V   &     K4V   &     G2V     &  4.76    &  0.26782       & \phn206\phm{.}\phn\phn\phn  & 78.39 $\pm$ 1.03 & 3 &          358 &241 & 753   & 0.600 V  & 0.49 V  & y & 1,2,5,60 \\
 148121 &  V1055~Sco  & \multicolumn{3}{c}{ G3V  }      &  8.64    &  0.36367       &  ---                        & 11.45 $\pm$ 1.25 & 3 &            4 &  3 &  12   & 0.250 V  & 0.25 V  & n & 1,2,61 \\
 157482 &  V819~Her   & F2V   &     F8V   &   G7III-IV  &  5.57    &  2.22964       & \phn\phn\phn5.530           & 15.53 $\pm$ 1.16 & 3 &           63 & 34 &  34   & 0.085 V  & 0.05 V  & y & 1,2,62,63 \\
 162724 &  V906~Sco   & B9V   &     B9V   &     B9      &  5.96    &  2.78595       & \phn\phn34.71\phn           &  3.23 $\pm$ 0.83 & 3 &            6 &  1 &  14   & 0.270 V  & 0.25 V  & Y & 1,2,64 \\
 163151 &  V2388~Oph  & \multicolumn{2}{c}{F3V}&  F     &  6.26    &  0.80230       & \phn\phn\phn8.925           & 14.72 $\pm$ 0.81 & 3 &           21 & 10 &  35   & 0.280 Hp & 0.23 Hp & y & 1,2,65,66 \\
 163708 &  V1647~Sgr  & A1V  &      A2V    &    F0-1V   &  6.8\phn &  3.28279       & 1219.7\phn\phn              &  8.70 $\pm$ 1.40 & 3 &            9 & 11 &  15   & 0.630 V  & 0.49 V  & Y & 1,2,67 \\
 165590 &  V772~Her   & G1V  &      K6V    &   K7V+M0V  &  7.10    &  0.87950       & \phn\phn20.08\phn           & 26.51 $\pm$ 1.35 & 5 &           29 &  0 & 256   & 0.100 V  &         & y & 1,2,68,69 \\
 184242 &  V2083~Cyg  & \multicolumn{3}{c}{ A3 }        &  6.88    &  1.86749       & \phn372\phm{.}\phn\phn\phn  &  3.98 $\pm$ 0.79 & 3 &            5 &  2 &  58   & 0.240 Hp & 0.24 Hp & y & 1,2,7,9 \\
 185936 &  QS~Aql     & B5V  &      F3     &   B4       &  5.99    &  2.51331       & \phn\phn61.72\phn           &  1.98 $\pm$ 0.82 & 3 &           21 &  3 &  76   & 0.130 V  & 0.04 V  & y & 1,2,70,71 \\
 187949 &  V505~Sgr   & A2V  &    G5IV    &   F7-8V     &  6.49    &  1.18287       & \phn\phn60.14\phn           &  8.58 $\pm$ 1.38 & 3 &          316 &  6 &  17   & 1.050 V  & 0.17 V  & y & 1,2,72,73 \\
 195434 &  MR~Del     & K2   &     K6     &    K2       & 11.01    &  0.52169       & 1996.6\phn\phn              & 22.53 $\pm$ 5.13 & 3 &           16 & 11 &  37   & 0.310 V  & 0.17 V  & Y & 1,2,17 \\
 197433 &  VW~Cep     & K1   &     G5     &    K3       &  7.3\phn &  0.27832       & \phn\phn29.79\phn           & 36.16 $\pm$ 0.97 & 3 &         1093 &821 &  17   & 0.450 V  & 0.33 V  & y & 1,2,13,74 \\
 201427 &  BR~Ind     &  \multicolumn{3}{c}{ F8V }      &  7.1\phn &  0.89277       & \phn167\phm{.}\phn\phn\phn  & 20.47 $\pm$ 2.08 & 3 &            1 &  0 &  37   & 0.140 Hp &         & y & 1,2,69
 \\ \hline
 \end{tabular}
 \caption{\scriptsize Columns 3 - 5 give the individual spectral types of primary, secondary and tertiary
components, $V$ denotes the magnitude in $V$ filter, $P$ stands for the orbital period of the
eclipsing pair, $p_3$ gives the period of the third body, $\pi$ quotes the Hipparcos parallax,
column 10 gives the total number of known components in the system, columns 11 and 12 stand for the
number of times of primary and secondary minima observed (including our new ones), $M$ denote the
number of astrometric observations, columns 14 and 15 give the depths of primary and secondary
minima. In column 16 "y/n" indicates if there is (or not) an orbital solution for the system. The
upper case `Y' denotes the orbital solution presented here for the first time. Column 17 lists the
references for the particular system, respectively. The orbital periods $p_3$ are adopted from the
published papers, or as determined in the present analysis., References: (1) \cite{GCVS2004}; (2)
\cite{HIP}; (3) \cite{1983IBVSBrettman}; (4) \cite{Griffin1999}; (5) \cite{Soderhjelm1999V640Cas};
(6) \cite{Olevic2002V348And}; (7) \cite{Seymour2002V2083Cyg}; (8) \cite{Abt1981}; (9)
\cite{Abt1985}; (10) \cite{BBSAG125}; (11) \cite{Andersen83}; (12) \cite{Clausen}; (13)
\cite{ZascheWolf}; (14) \cite{1969Newburg}; (15) \cite{1999Mason}; (16) \cite{2001BBScl}; (17)
\cite{Cutispoto1997}; (18) \cite{Appenzeller1967}; (19) \cite{Sch2007}; (20)
\cite{Popovic1995V773Cas}; (21) \cite{Chambliss1992}; (22) \cite{Zaera85}; (23)
\cite{Pan1993Algol}; (24) \cite{Lestrade1993}; (25) \cite{Gimenez1994}; (26) \cite{Rucinski2007};
(27) \cite{Grenier1999}; (28) \cite{Balega1999}; (29) \cite{2002DelOri}; (30)
\cite{Andersen1990V1031Ori}; (31) \cite{Leeuwen1997TauCMa}; (32) \cite{Leung1978}; (33)
\cite{Docobo2008}; (34) \cite{Ginestet2002}; (35) \cite{Malkov2006}; (36) \cite{Wolf2008}; (37)
\cite{Andersen1984}; (38) \cite{Docobo2007LOHya}; (39) \cite{Otero2000}; (40) \cite{Argyle2002};
(41) \cite{Strohmeier1959AN}; (42) \cite{ObjevDILyn1998}; (43) \cite{Tokovinin2006DILyn}; (44)
\cite{Tokovinin1998AstL}; (45) \cite{Chamberlin1957}; (46) \cite{Roberts2005}; (47)
\cite{Walborn1973}; (48) \cite{Murdoch1995}; (49) \cite{Parsons2004}; (50) \cite{Popper1986}; (51)
\cite{Aristidi1999DNUMa}; (52) \cite{VVCrv2008}; (53) \cite{Malaroda1975}; (54)
\cite{Finsen1964V831Cen}; (55) \cite{Edwards1976}; (56) \cite{Walker1985}; (57) \cite{Lu2001}; (58)
\cite{Seymour2001ETBoo}; (59) \cite{ETBoo2002BaltA}; (60) \cite{Docobo2006}; (61) \cite{Houk1982};
(62) \cite{vanHamme1994}; (63) \cite{Muterspaugh2006}; (64) \cite{V906Sco1997}; (65)
\cite{Yakut2004}; (66) \cite{Hartkopf1996}; (67) \cite{Andersen1985}; (68) \cite{Reglero1991}; (69)
\cite{Fekel1994}; (70) \cite{Holmgren1987}; (71) \cite{Zasche2008}; (72) \cite{Tomkin1992}; (73)
\cite{Cvetkovic2008}; (74) \cite{Kaszas1998}; (75) \cite{Houk1978}; (76) \cite{Duerbeck2007};}
 \label{Table1}
 \end{sidewaystable*}}}

 \normalsize

\section{The Catalog} \label{Catal}

The selection criteria used during the compilation of this catalog were as follows:
\begin{itemize}
 \item the existence of more than 10 astrometric measurements,
 \item a variation of position angles of the astrometric observations of more than 10 degrees, and
 \item the presence of an eclipsing binary as one of its components.
\end{itemize}

A total of 44 systems which met these criteria were found; these are presented in Table
\ref{Table1} and commented upon in more detail in the following subsections. In a few cases, it was
rather difficult to identify which component in the wide system comprised the eclipsing binary.
Other interesting systems, which contain an EB as one component but which do not satisfy the other
selection criteria, were also included in subsection \ref{Other}. In subsection \ref{Special} we
discuss those systems which were difficult to classify. These include, for example, systems where
the eclipsing nature is questionable, etc. In many systems the only information about their
distance is that by Hipparcos \citep{HIP}, see column 9 in Table \ref{Table1}. However, the value
of Hipparcos distance could be affected by relatively large error due to the fact the star is not a
single target, but a binary.

The entire catalog is also available
online\footnote{http://sirrah.troja.mff.cuni.cz/$\sim$zasche/Catalog.html}. This web site will be
updated as new observations become available and as new systems meeting our selection criteria are
discovered.

Notes on individual systems follow.

\subsection{V640 Cas}
\noindent \objectname{V640 Cas} (HD~123, HR~5, HIP~518, STF~3062AB) is listed as an eclipsing
binary. Only two photometric decreases were observed \citep{1983IBVSBrettman}, and no photometric
analysis has been published. This system was observed over 16 nights from July 2007 to November
2008 in $B$, $V$, and $R$ filters, but no minima were observed. From our new photometric
observations we can conclude that there is no variability above a level of 0.02~mag with periods in
interval 1.02 to 1.1 days. \cite{Griffin1999} discussed the plausibility of the photometric
variations observed by \cite{1983IBVSBrettman}, and also noted that no such variability is
observable from the Hipparcos satellite (see \citealt{HIP}), concluding that the system is not
variable at all. Therefore, the classification of V640 Cas as eclipsing system is questionable.
However, the system could be more complicated and due to the presence of the third body the
eclipses could turn on and off, similarly to V907~Sco \citep{Lacy1999}, SS~Lac \citep{Torres2001},
or V699~Cyg \citep{Lippky1994}. The visual orbit is well observed, with some 572 data points
obtained over 170 years. \cite{Soderhjelm1999V640Cas} computed the most recently published orbital
parameters, finding a period of about 107~yr and an angular semimajor axis of about 1.4\arcs.

\subsection{V348 And}  
\noindent \objectname{V348 And} (HD~1082, HIP~1233, A~1256AB) is an Algol-type EB. Its situation is
similar to that of V640 Cas, as neither times of minima nor a photometric analysis have been found
in the literature. The single available time of minimum was based on Hipparcos measurements, but is
only poorly covered. Our new observations of the system, obtained during 19 nights, were summarized
in \cite{IBVSV348And}. The visual orbit is defined by data  obtained over 93 years and covering the
range from 4 to 223 degrees in position angle. Two rather different orbital solutions have been
published: \cite{Olevic2002V348And} derived values $p_3 = 138$~yr and $a=150$~mas, while
\cite{Seymour2002V2083Cyg} found a period of about 330~yr and a semimajor axis 290~mas. Both
solutions appear to fit the limited arc of observations equally well. The mass sums derived from
these two different fits are not able to decide which solution is the more plausible, because the
light-curve analysis has not yet been performed and the individual spectral types are not known. If
we assume the spectral types of each of the components to be the same as the spectrum of the system
as a whole (B9IV), the resulting mass sum should be much higher than predicted by either orbital
solution -- $2.8~M_\odot$ for \cite{Olevic2002V348And} and $3.4~M_\odot$ for
\cite{Seymour2002V2083Cyg}.

\subsection{V355 And}  
\noindent \objectname{V355 And} (HD~4134, HIP~3454, STF~52AB) is also an Algol-type EB, with a
spectral classification of the whole system as F6V. The only published complete light curve is that
of \cite{BBSAG125}, who also determined 9 times of minima (unpublished, see the author's web
site)\footnote{http://www.student.oulu.fi/$\sim$ktikkane/AST/V355AND.html}. Observations of the
visual pair have been obtained during the past 170 years, but cover a range in position angles of
only about 20$^\circ$. No visual orbit has yet been computed, but the orbital coverage to date
suggests a period of order 3,000 years.

\subsection{$\zeta$ Phe}
\noindent $\zeta$~Phe (\objectname{HD 6882}, HR 338, HIP 5348) is an eccentric EB of Algol type,
its spectral types were determined as B6V + B8V (according to \citealt{Andersen83}). It is a visual
triple and double--lined spectroscopic binary. A detailed analysis using a combined solution of
astrometry and times of minima was presented by ZW. This analysis yields a period for the visual
orbit of about 221~yr. The mass of the predicted third body was derived to be $M_3 = 1.73~$\Mo,
which is in excellent agreement with earlier photometric analyses by \cite{Clausen} and
\cite{Andersen83}. In addition to the long-term variation seen in the $O-C$ diagram caused by the
light-time effect, apsidal motion is also detectable, with a period of $\sim$60~years.

\subsection{BB Scl} \noindent \objectname{BB Scl} (HD~9770, HIP~7372, GJ~60) is an
eclipsing binary and also the B component of a visual triple; we therefore deal with a quadruple
system. Orbits of both the visual pairs have been derived. The wider AB-C pair (1.42\arcs) revolves
on its 112~yr orbit \citep{1969Newburg}, while the closer A-B pair (0.17\arcs) with period of
4.6~yr was derived by \cite{1999Mason}. The most detailed analysis is that by \cite{2001BBScl}, who
analyzed this chromospherically active eclipsing binary using spectroscopic and photometric
techniques.

\subsection{V773 Cas}  
\noindent \objectname{V773 Cas} (HD~10543, HR~499, HIP~8115, BU~870AB) is an Algol-type EB. One
time of minimum light was derived from the Hipparcos observations, and three new minima were
observed by the authors (see Table \ref{Minima}). The astrometry covers about 80$^\circ$ in
position angle, from observations obtained during the last 120 years. The most recent preliminary
visual orbit calculation by \cite{Popovic1995V773Cas} derives a period of about 304~yr and a
semimajor axis about 1\arcs.

\subsection{AA Cet}  
\noindent \objectname{AA Cet} (HD~12180, HIP~9258, ADS~1581~A) is a W~UMa-type EB. There have been
more than 200 times of minima obtained during the last 40 years, but no significant LITE variation
has been detected. Astrometric observations of the visual pair have shown no significant change
since its discovery by William Herschel in 1782; hence no orbital solution has been attempted. One
new minimum of light was observed at Athens Observatory (see Table \ref{Minima}). Recently,
\cite{Duerbeck2007} discovered the visual component to be also a binary.

\begin{longtable}{p{0.6in} ccccc}
 \tablecaption{New minima timings of selected systems based on CCD and photoelectric observations,
 the Kwee-van Woerden (\citeyear{Kwee}) method was used. \label{Minima}}
 \tablewidth{\textwidth}
 \tablehead{\colhead{Star} & \colhead{HJD-2400000} & \colhead{Error} & \colhead{Type} & \colhead{Filter} &
 \colhead{Obs.}} \\[-2mm]
 V773 Cas & 54507.4103 & 0.0005 &   Pri   &   R   & [1] \\
 V773 Cas & 54533.2864 & 0.0007 &   Pri   &   R   & [1] \\
 V773 Cas & 54776.4933 & 0.0003 &   Pri   &   I   & [4] \\
   AA Cet & 53687.4389 & 0.0002 &   Pri   &   R   & [2] \\
 V559 Cas & 54433.3446 & 0.0004 &   Sec   &  BVR  & [1] \\
 V559 Cas & 54505.2635 & 0.0004 &   Pri   &   R   & [1] \\
 V559 Cas & 54535.2957 & 0.0007 &   Pri   &   R   & [1] \\
 V559 Cas & 54738.4053 & 0.0009 &   Sec   &  BVRI & [4] \\
 V559 Cas & 54810.3245 & 0.0025 &   Pri   &  BVRI & [4] \\
 V592 Per & 54432.4787 & 0.0003 &   Sec   &   R   & [1] \\
 V592 Per & 54491.5237 & 0.0003 &   Pri   &   R   & [1] \\
 V592 Per & 54517.2911 & 0.0003 &   Pri   &   R   & [1] \\
 V592 Per & 54523.3741 & 0.0005 &   Sec   &   R   & [1] \\
 V592 Per & 54774.5904 & 0.0003 &   Sec   &   R   & [4] \\
 V592 Per & 54798.5679 & 0.0003 &   Pri   &   R   & [4] \\
 V592 Per & 54826.4795 & 0.0005 &   Pri   &   R   & [4] \\
V1031 Ori & 54831.3964 & 0.0003 &   Sec   &   R   & [4] \\
V1031 Ori & 54860.3377 & 0.0004 &   Pri   &   R   & [4] \\
 V635 Mon & 54539.2564 & 0.0014 &   Pri   &   BV  & [2] \\  
 V635 Mon & 54857.4412 & 0.0019 &   Pri   &   R   & [4] \\
   AC UMa & 54620.4612 & 0.0003 &   Pri   &   R   & [3] \\
   DI Lyn & 54585.3703 & 0.0015 &   Sec   &   R   & [1] \\
   DI Lyn & 54591.2478 & 0.0020 &   Pri   &   R   & [1] \\
   DI Lyn & 54912.4254 & 0.0017 &   Pri   &  VRI  & [4] \\
   DI Lyn & 54933.4489 & 0.0025 &   Sec   &  BVR  & [1] \\
   TX Leo & 54595.4140 & 0.0021 &   Pri   &   R   & [4] \\
   DN UMa & 52381.7956 & 0.0003 &   Sec   &   BV  & [5] \\
   DN UMa & 54508.4873 & 0.0012 &   Sec   &   R   & [1] \\
   DN UMa & 54521.4630 & 0.0009 &   Pri   &   R   & [1] \\
   DN UMa & 54834.6816 & 0.0012 &   Pri   &   R   & [1] \\
   HT Vir & 54539.4315 & 0.0003 &   Pri   &   VR  & [2] \\  
   HT Vir & 54539.6368 & 0.0005 &   Sec   &   VR  & [2] \\
   ET Boo & 54524.4746 & 0.0002 &   Pri   &  VRI  & [3] \\
    i Boo & 54499.4495 & 0.0003 &   Pri   &   R   & [1] \\
    i Boo & 54499.5855 & 0.0001 &   Sec   &   R   & [1] \\
    i Boo & 54499.7179 & 0.0001 &   Pri   &   R   & [1] \\
V1055 Sco & 53090.1272 & 0.0025 &   Pri   &   V   & [6] \\
V1055 Sco & 53090.3076 & 0.0031 &   Sec   &   V   & [6] \\
 V819 Her & 54564.3794 & 0.0040 &   Pri   &   R   & [1] \\
 V819 Her & 54585.5559 & 0.0020 &   Sec   &   R   & [1] \\
 V819 Her & 54594.4766 & 0.0010 &   Sec   &   R   & [4] \\
 V819 Her & 54623.4631 & 0.0010 &   Sec   &  BVR  & [1] \\
 V819 Her & 54642.4133 & 0.0030 &   Pri   &  BVR  & [4] \\
 V819 Her & 54738.2917 & 0.0021 &   Pri   &   R   & [1] \\
V2388 Oph & 54534.6075 & 0.0030 &   Sec   &   R   & [2] \\ 
V2388 Oph & 54583.5470 & 0.0003 &   Sec   &   R   & [4] \\
V2388 Oph & 54593.5738 & 0.0002 &   Pri   &   R   & [4] \\
V2388 Oph & 54614.4372 & 0.0006 &   Pri   &   R   & [4] \\
V2388 Oph & 54620.4524 & 0.0005 &   Sec   &   R   & [4] \\
V2388 Oph & 54675.4117 & 0.0004 &   Pri   &   R   & [1] \\
V2388 Oph & 54718.3351 & 0.0004 &   Sec   &   R   & [1] \\
 V772 Her & 54540.5504 & 0.0021 &   Pri   &   R   & [4] \\
 V772 Her & 54584.5230 & 0.0008 &   Pri   &   R   & [4] \\
 V772 Her & 54628.5031 & 0.0009 &   Pri   &   R   & [1] \\
 V772 Her & 54713.8093 & 0.0010 &   Pri   &   R   & [1] \\
V2083 Cyg & 54583.5507 & 0.0003 &   Pri   &   R   & [1] \\
V2083 Cyg & 54598.4903 & 0.0007 &   Pri   &  BVR  & [1] \\
V2083 Cyg & 54683.4603 & 0.0008 &   Sec   &  BVR  & [1] \\
V2083 Cyg & 54684.3951 & 0.0006 &   Pri   &  BVR  & [1] \\
V2083 Cyg & 54698.4014 & 0.0005 &   Sec   &  BVR  & [1] \\
V2083 Cyg & 54712.4065 & 0.0005 &   Pri   &  BVR  & [1] \\
   QS Aql & 54309.4160 & 0.0030 &   Pri   &   R   & [1] \\
   QS Aql & 54383.5446 & 0.0005 &   Sec   &   R   & [1] \\
   QS Aql & 54696.4583 & 0.0020 &   Pri   &   R   & [4] \\
   QS Aql & 54726.6177 & 0.0005 &   Pri   &   R   & [1] \\
   QS Aql & 54726.6177 & 0.0006 &   Pri   &   R   & [1] \\
 V505 Sgr & 52837.4766 & 0.0030 &   Pri   & BVRI  & [2] \\
 V505 Sgr & 52843.3892 & 0.0029 &   Pri   &  VRI  & [2] \\
 V505 Sgr & 53263.3029 & 0.0002 &   Pri   &   R   & [2] \\
 V505 Sgr & 54267.5470 & 0.0002 &   Pri   &   VI  & [2] \\
 V505 Sgr & 54270.5016 & 0.0040 &   Sec   &   I   & [2] \\
 V505 Sgr & 54648.4260 & 0.0005 &   Pri   &   R   & [4] \\
 V505 Sgr & 54655.5233 & 0.0003 &   Pri   &   R   & [4] \\
 V505 Sgr & 54658.4817 & 0.0005 &   Sec   &   R   & [4] \\
 V505 Sgr & 54706.3869 & 0.0002 &   Pri   &   R   & [4] \\
   MR Del & 54278.4913 & 0.0005 &   Sec   &   R   & [3] \\
   MR Del & 54676.5415 & 0.0010 &   Sec   &   R   & [4] \\
   MR Del & 54682.5395 & 0.0010 &   Pri   &  BVR  & [4] \\
   MR Del & 54706.5374 & 0.0003 &   Pri   &   R   & [4] \\
   VW Cep & 54521.2597 & 0.0002 &   Sec   &   R   & [1] \\
   VW Cep & 54522.3723 & 0.0003 &   Sec   &   R   & [1] \\
   VW Cep & 54522.5128 & 0.0003 &   Pri   &   R   & [1] \\
   VW Cep & 54536.2892 & 0.0003 &   Sec   &   R   & [1] \\
   VW Cep & 54536.4282 & 0.0002 &   Pri   &   R   & [1] \\
   VW Cep & 54536.5670 & 0.0002 &   Sec   &   R   & [1] \\ \hline 
\end{longtable}
\tablecomments{\scriptsize [Obs.:] [1] - P.\ Svoboda, Brno; [2] - Athens Observatory; [3] -
Ond\v{r}ejov Observatory; [4] - R.Uhla\v{r}, J\'{\i}lov\'e u Prahy; [5] - San Pedro M\'artir
Observatory; [6] - OMC INTEGRAL satellite }

\subsection{V559 Cas}  
\noindent \objectname{V559 Cas} (HD~14817, HIP~11318, STF~257AB) is an Algol-type eclipsing and
spectroscopic binary. There have been only 14 observed times of minima since 1971, including our
latest ones (see Table \ref{Minima}). Due to its very long orbital period (about 836~yr, according
to \citealt{Zaera85}), only about 100$^\circ$ of the orbit has been observed since 1830. Periastron
passage occurred in 1932 and was well-covered; regrettably, no minima times were determined during
that era.

\subsection{$\beta$ Per}  
\noindent Algol ($\beta$~Per, \objectname{HD 19356}, HR~936, HIP~14576, LAB~2Aa,Ab) is the
well-known prototype of this class of binaries. With its 2.12~mag in \emph{V} it is the second
brightest system in the catalog. The time of minimum brightness was first measured by Montanari on
8 November 1670 (although the variability of the ``Demon Star" had been known from much earlier
times). The current set of times of minima is very large, with about 1400 observations covering the
past three centuries, and photoelectric measurements dating as far back as 1910
(\citealt{Stebbins1910}). In spite of this, a detailed description of period variations in the
$O-C$ diagram is still missing. The system is rather complicated, but the distant component
(orbital period $\sim$1.8~yr) originally discovered on the basis of  radial velocity variations,
was first resolved by speckle interferometry in 1973 \cite{Labeyrie1974}. The orbit of this
component is now well established ($a=94.6$~mas and $e=0.23$, according to \citealt{Pan1993Algol}).
Several distant companions are listed in the WDS, but these are probably optical.

\subsection{AG Per}  
\noindent \objectname{AG Per} (HD~25833, HIP~19201, STT~71AB) is an Algol-type EB. Over 100 times
of minima, obtained from the 1920's up to the present, have been collected from the published
literature. AG~Per is one of the most typical apsidal--motion systems, and has been analyzed
several times (see e.g. \citealt{Wolf2006AGPer}). Data are sufficient for combining the apsidal
motion and LITE into one joint solution (similar to the $\zeta$~Phe case). Precise light curves
have also been measured and analyzed (see \citealt{Woodward1987}). On the other hand, although
astrometric observations have been obtained 39 times since 1846, the change in position angle has
been too small to permit an orbital solution.

\begin{figure}
 \plotone{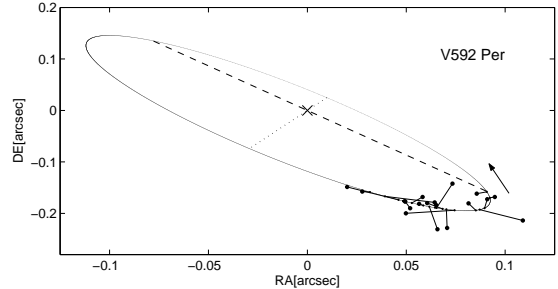}
 \caption{Relative orbit of V592~Per on the plane of the sky. Individual measurements are
 plotted as dots, and straight ``O-C" lines connect the these observations to their predicted
 positions on the fitted orbit. The cross indicates the position of the eclipsing binary on the
 sky, the dotted line represents the line of apsides and the dashed line the line of nodes.}
\label{V592Per}
\vskip 0.1in ~
\end{figure}

\subsection{V592 Per}  
\noindent \objectname{V592 Per} (HD~29911, HIP~22050, COU~1524) is a $\beta$-Lyrae EB system. There
have been 25 times of minima published to date, six of them observed for this paper (see Table
\ref{Minima}). Astrometric data cover only about 20$^\circ$, with 19 data points obtained over 28
years. A preliminary orbit has been computed (using standard methods, see e.g.
\citealt{1973Batten}), yielding a period of about 115~yr and a semimajor axis of about 220~mas. The
derived elements are given in Table \ref{Orbits} and plotted in Fig.\ \ref{V592Per}, together with
all observations used in the solution. Due to the fact the orbit is only a preliminary one, the
resulting derived mass of the system is unreliable. In this case the mass derived from the visual
orbit (about 6\Mo) is considerably higher than would be expected according to the estimated
spectral types. New observations are needed.

\begin{deluxetable*}{ c l | c c c c c c c | c }
  \tabletypesize{\tiny}
 \tablecaption{Orbital elements for the newly derived solutions and the total masses of the systems according to these orbital solutions. \label{Orbits}}
 \tablewidth{\textwidth}
 \tablehead{ \\[2mm]
 \colhead{WDS designation} & \colhead{Variable} & \colhead{Period} & \colhead{Epoch of} & \colhead{Semimajor} & \colhead{Inclination} & \colhead{Longitude} & \colhead{Eccentricity} & \colhead{Longitude} & \colhead{Total}
 \\[1mm]
 \colhead{ } &    \colhead{star} &   \colhead{ } &  \colhead{periastron} &  \colhead{axis} &  \colhead{ } &  \colhead{of periastron} &  \colhead{ } &  \colhead{of node} &  \colhead{mass}
 \\[1mm]
 \colhead{$\alpha \delta$(2000)} & \colhead{designation} & \colhead{$p_3$[yr]}&   \colhead{$T_0$[yr]} & \colhead{$a$[\arcs]} & \colhead{$i$[deg]} & \colhead{$\omega$[deg]} & \colhead{$e$} & \colhead{$\Omega$[deg]} &
 \colhead{[\Mo]}
 }
 \\[2mm]
  04445+3953  &    V592~Per       &  \phn115.3\phn   &    1903.9    &   \phn0.223    &   \phn79.1  &   \phn98.5    &   0.498   &    210.0  & \phn\phn\phn6.21 \\[1mm]
   &  & \phn$\,\pm$7.1\phn &\phn\phn$\,\pm$6.4\phn&\phn$\,\pm$0.030\phn\phn&$\,\pm$8.4&$\,\pm$9.7&$\,\pm$0.006\phn\phn&\phn$\,\pm$9.0\phn  & \phn$\,\pm$4.08 \\[1mm]
  05320-0018  &  $\delta$~Ori~A   &  \phn704.8\phn   &    1972.8    &   \phn0.970    &   \phn96.9  &   \phn95.2    &   0.914   &    325.3  & \phn\phn40.7\phn \\[1mm]
   &  & $\pm$127.7$\;$\phn &$\pm$145.2$\;$&\phn$\,\pm$0.331\phn\phn&$\pm$13.8$\;$&$\,\pm$9.6&$\,\pm$0.165\phn\phn&$\pm$17.3$\:$ &  $\,\pm$32.2\phn \\[1mm]
  05474-1032  &   V1031~Ori       &  \phn\phn92.66   &    1942.6    &   \phn0.176    &   \phn76.3  &   \phn\phn0.3 &   0.001   &    111.9  & \phn\phn\phn5.11 \\[1mm]
   &  & \phn$\,\pm$6.61 &\phn$\,\pm$7.8&\phn$\,\pm$0.007\phn\phn&\phn$\pm$8.2$\;$&$\pm$28.5$\:$&$\,\pm$0.001\phn\phn&\phn$\pm$9.4$\;$  &   \phn$\,\pm$1.13\\[1mm]
  08558+6458  &      AC~UMa       &  1199.2\phn      &    1349.0    &   30.15\phn    &   106.5     &   102.7       &   0.790   &    \phn62.9  & $5\!\cdot\!10^6$ \\[1mm]
   &  & $\pm$341.8$\;$\phn &$\pm$280.5$\;$&\phn$\,\pm$0.52\phn\phn\phn&\phn$\pm$7.8$\;$&$\:\pm$8.9&$\,\pm$0.047\phn\phn&\phn$\pm$6.8$\;$ & $\pm4\!\cdot\!10^6$\\[1mm]
  11383-6322  &    V871~Cen       &  1715.8\phn      &   \phn909.2  &   \phn1.316    &   \phn81.9  &   \phn73.8    &   0.730   &    234.8 & 3390 \\[1mm]
   &  & $\pm$285.4$\;$\phn &$\pm$230.1$\;$&\phn$\,\pm$0.445\phn\phn&$\pm$13.0$\;$&$\pm$14.8$\:$&$\,\pm$0.106\phn\phn&\phn$\pm$8.8$\;$ & $\!\!\pm$2860\phn \\[1mm]
  17539-3445  &    V906~Sco       &  \phn\phn34.71   &    1974.1    &   \phn0.172    &   \phn80.5  &   183.0       &   0.001   &    103.0 & \phn125.2 \\[1mm]
   &  & \phn$\,\pm$1.13 &\phn$\,\pm$1.2&\phn$\,\pm$0.015\phn\phn&\phn$\pm$6.8$\;$&\phn$\pm$9.7$\;$&$\,\pm$0.002\phn\phn&\phn$\pm$9.7$\;$ & \phn$\!\!\pm$46.7\\[1mm]
  17592-3656  &   V1647~Sgr       &  1219.7\phn      &    1778.4    &   14.36\phn    &   \phn94.7  &   269.4       &   0.841   &    113.8 & 3021 \\[1mm]
   &  & $\:\pm$83.4\phn &$\:\pm$76.2&$\:\pm$6.08\phn\phn&\phn$\pm$8.5$\;$&$\pm$21.2$\;$&$\,\pm$0.027\phn\phn& $\pm$12.7$\;$  & $\!\!\pm$468 \\[1mm]
  20312+0513  &      MR~Del       &  1996.6\phn      &    1945.9    &   \phn3.763    &   \phn74.2  &   \phn\phn0.0 &   0.493   &  \phn61.2 & \phn\phn1.17 \\[1mm]
   &  & $\pm$170.5$\;$\phn &$\pm$130.3$\;$&\phn$\,\pm$0.156\phn\phn&\phn$\pm$8.4$\;$&\phn$\pm$7.4$\;$&$\,\pm$0.051\phn\phn&\phn$\pm$8.0$\;$ &
   \phn$\!\!\pm$0.40 \\[1mm]
\end{deluxetable*}

\subsection{$\eta$ Ori}  
\noindent $\eta$~Ori (\objectname{HD 35411}, HR~1788, HIP~25281, MCA~18Aa,Ab + DA~5AB) is a
quadruple system with three resolvable components. The primary is an eclipsing and also a
double-lined spectroscopic triple (periods 8~d and 9.2~yr). One of the components also has a
pulsation period of $\sim$8~hours. There has been only one published time of minimum, derived from
the Hipparcos observations. The orbit of the interferometric pair (MCA~18Aa,Ab) derived by
\cite{Balega1999} finds an orbital period of about 9.44~yr, in reasonable agreement with the
spectroscopic period.

\begin{figure}
  \plotone{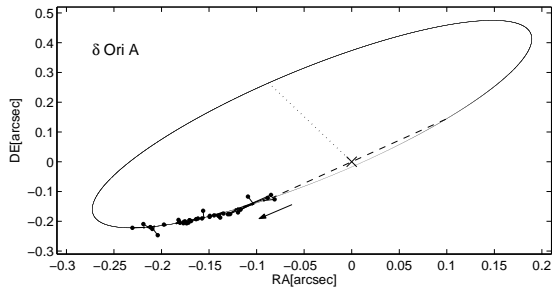}
  \caption{Relative orbit of $\delta$~Ori~A on the plane of the sky.} \label{DelOri}
\end{figure}

\subsection{$\delta$ Ori A} 
\noindent $\delta$~Ori~A (\objectname{HD 36486}, HR~1852, HIP~25930, HEI~42Aa,Ab) is a massive
eclipsing binary with an orbital period 5.7 day. Only 9 times of minima were found in the
literature, and these minima do not show any significant LITE variation (more probably apsidal
motion). On the other hand, there is a fair amount of motion on the plane of the sky seen in the 38
measurements obtained since the visual pair was first observed in 1978. The preliminary orbit
listed in Table \ref{Orbits} and shown in Fig.\ \ref{DelOri} predicts a period of about 705~yr and
a semimajor axis of 1\arcs. From the orbital parameters, the total mass of the system is about
40\Mo. For a detailed discussion about the masses of the binary components, see \cite{2002DelOri}.
There appears to be a problem with the derived masses of primary and secondary, which are
substantially below the expected masses for stars of their luminosity. On the other hand, the
evolutionary tracks of stars with the measured values of [$\log T_{eff}, \log L$] predict much
higher masses for both components. If we accept the masses derived by \citeauthor{2002DelOri}, the
mass of the third component should be about 20\Mo. Single stars of such a mass should be observable
via a UV flux contribution, which has not happened. A plausible conclusion is that the distant
component is probably also a binary.

\subsection{V1031 Ori}   
\begin{figure}
  \plotone{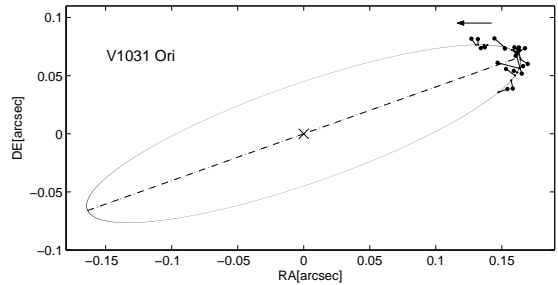}
  \caption{Relative orbit of V1031~Ori on the plane of the sky.} \label{V1031Ori}
\end{figure}

\noindent \objectname{V1031 Ori} (HD 38735, HR~2001, HIP 27341, MCA~22) is an Algol-type detached
system. Twelve times of minima have been found in the literature. A new circular orbit of the
interferometric binary is shown in Fig.\ \ref{V1031Ori}. Based on only 20 observations obtained
from 1980 to 1997, the solution is obviously very preliminary. The derived orbital period is about
93~yr and the semimajor axis about 0.18\arcs (see Table \ref{Orbits} for the parameters). From our
new orbital solution, we derive a mass of $M_{123} = (5.1 \pm 1.1)$\Mo~ for all three components.

Based on the RV measurements and a few speckle observations, \cite{Andersen1990V1031Ori} concluded
that the orbit should be much larger, with a period about 3700~yr. Third-component lines were
observed in the spectra of V1031~Ori and radial velocities for the 93~yr orbit would be much larger
than measured. \cite{Andersen1990V1031Ori} also derived the physical parameters of both eclipsing
components, resulting in $M_{12} = (4.76 \pm 0.04)$\Mo. However, they assumed that the wide orbit
is coplanar with the eclipsing binary orbit, which is not necessary true in multiple systems.
Because the orbital coverage is so poor, only further interferometric observations, as well as
precise radial velocity investigation will reveal the true nature of the system. New times of
minima are also needed.

\subsection{$\tau$ CMa}
\noindent $\tau$~CMa (\objectname{HD 57061}, HR~2782, HIP~35415, FIN~313Aa,Ab) is the brightest
star in the open cluster NGC~2362. It is a $\beta$~Lyrae-type EB, with a period about 1.28~d.
$\tau$~CMa is also a spectroscopic binary with an orbital period of about 154.9~days; the EB is
probably the visual secondary \citep{Stickland1998}. This interesting system therefore contains
both the longest period spectroscopic binary and the shortest period eclipsing binary known among
the O-type stars. The system was precisely analyzed by \cite{Leeuwen1997TauCMa}. This triple system
is one member of the visual binary FIN~313Aa,Ab, which has been measured astrometrically 32 times
since 1951. The change in position angle is still quite small, however, so no orbital solution has
yet been attempted.

\subsection{YY Gem}
\noindent \objectname{YY Gem} (Castor~C, $\alpha$~Gem~C, HD~60178J, HIP~36850, STF~1110AB) is an
eclipsing binary and also the C component of the Castor multiple system. The three visual
components are all doubles: A and B are spectroscopic binaries, while C is the eclipsing binary
YY~Gem; we therefore are dealing with a sextuple system. (A fourth visual companion is also listed
in the WDS; however, the D component shows a very different proper motion and is likely an optical
rather than physical companion.) About 180 times of minima have been published for YY~Gem. With
1,341 observations covering nearly two centuries, the orbit of the AB pair ($p_3 = 467$~yr, $a =
6.8$\arcs) is well defined now. However, no significant change in position angle has been seen in
the orbit of component C around the AB pair, despite data spanning 180 years.

\subsection{V635 Mon}  
\noindent \objectname{V635 Mon} (HD~66094, HIP~39264, A~1580AB) is an Algol-type EB. It is
well-observed, with over 100 times of minima published to date; however these data points follow
the linear ephemeris without any indication of a LITE. A total of 24 astrometric measurements have
been made over the past century, spanning about 150$^{\circ}$ (although phase coverage was very
sparse through most of that time). The most recent analysis by \cite{Docobo2008} gives a period of
257~yrs, with a semimajor axis of about 313~mas.

\subsection{NO Pup}  
\noindent \objectname{NO Pup} (HD~71487, HR~3327, HIP~41361, B~1605Ba,Bb) is an eccentric eclipsing
binary of Algol type. There have been 15 times of minima observed and the apsidal motion of the
system has been studied a few times, resulting in an apsidal period of about 37~yr
\citep{Gimen1986}. Altogether there are 4 visible components in this multiple system, with the
eclipsing system probably comprising the primary. \cite{Chambliss1992} has estimated that the close
Ba,Bb pair orbit with a period of about 32~yr, but no orbital analysis has yet been published. The
relative separation and angle between A and the pair comprising B has remained essentially
unchanged over the past 160 years. A fourth component was discovered in 1997, using adaptive
optics, by \cite{Tokovinin1999}, but has not yet been confirmed.

\subsection{VV Pyx}   
\noindent \objectname{VV Pyx} (V596~Pup, HD~71581, HR~3335, HIP~41475, B~2179AB) is an Algol-type
EB and also a double-lined spectroscopic binary. \cite{Andersen1984} analyzed the light curve and
also the RV curves of the system, deriving a precise set of physical parameters. There were 11
times of minima found in the literature (1976 -- 2005), but these data show very slow apsidal
motion (on the order of centuries). The visual orbit is also covered only very poorly, with 11
observations obtained over the course of 38~years showing a change in position angle of about
13$^\circ$; no orbital anaylsis is yet possible. \cite{Andersen1984} speculate that the visual
companion could also be a binary; speckle interferometric observations have ruled out any
companions with separations greater than $\sim$30~mas and magnitude difference less than about
3~mag, but only further precise observations could prove or disprove the existence of closer and/or
fainter companions.

\subsection{LO Hya}  
\noindent \objectname{LO Hya} (HD~71663, HR~3337, HIP~41564, A~551AB) is another Algol-type
eclipsing binary. A detailed analysis of this system was performed by \cite{Bakos1985LOHya}. Both
the A and B components are spectroscopic binaries, while the distant C component also appears to be
a physical companion. We therefore deal with at least a quintuple system. One spectroscopic
component is also the eclipsing binary LO~Hya, with an orbital period of about 2.5~day. Six times
of minima have been found in the literature. Some 69 astrometric measurements obtained during the
last century reveal a visual orbit with a 55-year period (see \citealt{Docobo2007LOHya}).

\subsection{$\delta$ Vel}  
\noindent $\delta$~Vel (\objectname{HD 74956}, HR~3485, HIP~42913, I~10AB) is an Algol-type
eclipsing binary classified as A1V spectral type; at $V = 1.95$~ mag it is the brightest system in
the catalog. The star was discovered to be photometrically variable in 1997 (see
\citealt{Otero2000} for details); the period of such variation is about 45 days. Altogether 10
times of minima were obtained, but these observations indicate very slow apsidal motion (on the
timescale of centuries). The visual A and B components orbit with a period of about 142 years and a
semi-major axis of about 2\arcs ~(according to \citealt{Alzner2000DelVel}). The whole system is,
however, more complicated, consisting of two proper motion pairs (with separations of 2\arcs and
6\arcs, respectively), separated in the sky by 72\arcs. The primary component was additionally
resolved into a 15~mas pair by long-baseline interferometry. We therefore appear to be dealing with
a system of at least 6 components. However, thanks to the interferometric observations and detailed
analysis by \cite{Kellerer2007DelVel}, the picture of the system has been simplified somewhat.
According to these authors, the two distant components C and D do not belong to the system.
Furthermore, the 45-day period interferometric orbit appears to correspond to that of the eclipsing
pair.

\begin{figure}
 \plotone{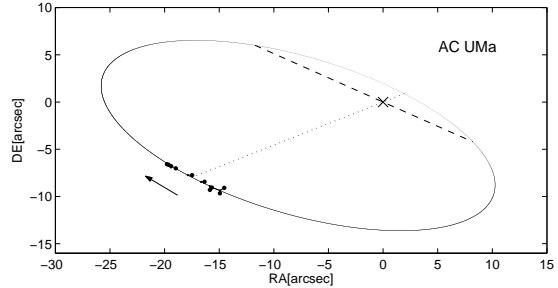}
 \caption{Relative orbit of AC~UMa on the plane of the sky.} \label{ACUMa}
\end{figure}

\subsection{AC UMa}  
\noindent \objectname{AC UMa} (ARG~21AB) is an Algol-type EB, and at $V = 10.3$~mag is one of the
faintest systems studied here. The orbital period is about 6.85~days. The 80 times of minima found
in the literature suggest a possible variation in orbital period on the timescale of decades, but
this finding is inconclusive and a larger data set is needed. The visual orbit shown in Fig.\
\ref{ACUMa} is based on only 10 observations obtained during the past 106 years. As one can see,
only a short arc of the orbit is covered by data, so one cannot derive the parameters of the orbit
precisely. This solution gives an extremely long period of about 1200~yr, although this value could
be even longer (see Table \ref{Orbits}). Another possible explanation is that the distant component
is only an optical companion and not physically associated with the EB system. The different proper
motions of the two visual components suggest this latter interpretation to be the more probable one
(see the Catalog of Rectilinear Elements\footnote{http://ad.usno.navy.mil/wds/lin1.html} for
details). Another argument that the component is probably not gravitationally bound with the EB is
the fact that the total mass computed from the visual orbit is unacceptably high, see Table
\ref{Orbits}.

\subsection{DI Lyn} 
\noindent \objectname{DI Lyn} (A~Hya, HD~82780, HR~3811, HIP~47053, COU~2084Aa,Ab) is an Algol-type
EB. There is only one time of minimum available in the literature, see \cite{ObjevDILyn1998};
however, four additional minima were measured for this paper (see Table \ref{Minima}). This
hierarchical system is rather complicated, and contains at least five physical components, see
\cite{Tokovinin2006DILyn}, component C seems to be only an optical one. Two components are
spectroscopic binaries (periods 28~d and 1.7~d), one of them is also the eclipsing binary DI~Lyn.
\cite{Tokovinin1997MSC} estimated the period of the close visual Aa,Ab pair at $\sim$64~yr.
However, any orbital solution has been published, and this estimation of period was based only on
the relative motion of these stars - only about 28$^{\circ}$ in 22 years - and the Kepler's third
law, so this may be an underestimate.

\subsection{TX Leo}  
\noindent \objectname{TX Leo} (HD~91636, HR~4148, HIP~51802, STF~1450AB) is an Algol-type EB and
its apparent brightness is about $V = 5.67$~mag. There have been 8 times of minima observed since
1930. The astrometric data set is much larger, about 140 measurements secured over the last 180
years, but the relative motion has been minimal.

\begin{figure}
  \plotone{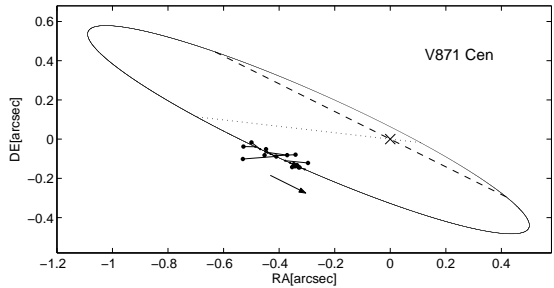}
  \caption{Relative orbit of V871~Cen on the plane of the sky.} \label{V871Cen}
\end{figure}

\subsection{V871 Cen}  
\noindent \objectname{V871 Cen} (HD~101205, HIP~56769, I~422AB) is a $\beta$~Lyrae-type EB. There
is a brief paper on the photometric observations of V871~Cen, together with one minimum time
derived (see \citealt{Mayer1992V871Cen}). Confirmation of the period of its photometric variability
is based on analysis of Hipparcos and ASAS data, see \cite{2007OEJV72}. The system includes 4
visual components; astrometric measurements of the closest pair secured during the last 104 years
reveal a change in position angle of only about 20$^\circ$, so our 1700-year period orbital
solution is not very convincing (see Table \ref{Orbits} and Fig.\ \ref{V871Cen} for details). Also
a derived total mass of the system more than 3000\Mo~indicates an unacceptable solution with the
current data.

\subsection{GT Mus}
\noindent \objectname{GT Mus} (HD~101379, HR~4492, HIP~56862, B~1705AB) is one component of a
quadruple system, as each member of this close visual pair is itself a close binary (see
\citealt{Murdoch1994GTMus}). One of them is a spectroscopic binary with period about 61~days, which
is also an RS~CVn-type binary, while the other one is an eclipsing binary with orbital period about
2.75~days. No minima have been published. Astrometric data obtained during 76~years and cover about
110$^\circ$ of the orbit. The estimated period of the AB pair is about 91~years
\citep{Parsons2004}. The visual pair has two faint wide companions; the physical/optical nature of
these are unknown.

\subsection{DN UMa}  
\noindent \objectname{DN UMa} (HD~103483, HR~4560, HIP~58112, A~1777AB) is another Algol-type EB,
which comprises the primary component of the visual quadruple system ADS 8347. The D component of
this system is over 1 arcminute in separation from the primary; however, the similarity in parallax
indicates the wide pair is probably physical. A detailed light-curve analysis was published by
\cite{Garcia1986} and an RV curve analysis by \cite{Popper1986}. There have been 16 observed times
of minima from 1979 until our recent measurements (see Table \ref{Minima}). The visual orbit is
defined reasonably well, with a century's worth of data covering nearly a full revolution. The most
recent orbit was by \cite{Aristidi1999DNUMa}, who derived $p_3 = 136.5$~yr and $a = 230$~mas.

\subsection{VV Crv}
\noindent \objectname{VV Crv} (HD~110317, HIP~61910, STF~1669AB) is a system consisting of two
spectroscopic binaries (periods 44.51~d and 1.46~d; see \citealt{VVCrv2008}). The eclipsing nature
of one of its components was discovered from Hipparcos data \citep{HIP}; this gave an orbital
period of about 3.14~days, suggesting that the system might be quintuple. Hipparcos data have also
provided the only time of minimum found thus far in the literature. Astrometric data have been
obtained over the past 180 years, but have described a change in position angle of about only
14$^\circ$. Due to this short arc, no orbit has been calculated; however, \cite{Tokovinin2006DILyn}
have estimated the period of AB pair to be about 4500~years.

\subsection{V831 Cen}  
\noindent \objectname{V831 Cen} (HD~114529, HR~4975, HIP~64425, SEE~170AB) is a $\beta$~Lyrae
system and also a spectroscopic binary, with an orbital period of about 0.64~d. The
only published minimum is that measured by the Hipparcos satellite. The close AB pair of this visual
quadruple system has been observed for over 100 years, with a single attempt at an orbital solution
by \cite{Finsen1964V831Cen} yielding $p_3 = 27$~yr and $a = 185$~mas. $O-C$ errors are quite large,
however; this is due primarily to the small separation of the AB pair, but perhaps also in part to the
presence of the nearby C component (separation $<$2$''$) further complicating the early measurements.

\subsection{HT Vir} 
\noindent \objectname{HT Vir} (HD~119931, HIP~67186, STF~1781) is a contact W~UMa system. Although
only a visual binary, the system is in fact quadruple, with three components visible in the
spectra. The spectroscopic single-lined binary (period 32.5 d) constitutes component A, while the
eclipsing binary (period 0.41 d) is the B component. Astrometric measurements have been made
regularly since Struve's discovery of the visual pair in 1830, defining the orbit quite precisely.
On the other hand, times of minima have been measured only a few times since discovery of the
eclipsing variable. A detailed analysis of this system, combining the angular position measurements
and period variation, was presented in ZW.

\subsection{ET Boo}
\noindent \objectname{ET Boo} (HIP~73346, COU~1760) is a 9$^{\rm th}$ magnitude $\beta$~Lyrae
eclipsing binary. The system has been found to be quadruple, according to \cite{Pribulla2006}.
There have been 31 times of minima published to date, including one new value published here (see
Table \ref{Minima}). It is also a close visual binary, discovered in 1978 (\citealt{Couteau1981}).
Astrometric measurements obtained between 1978 and 1999 have shown a change in position angle of
about 40$^\circ$; unfortunately it has not been observed in nearly a decade. A very preliminary
orbit was derived by \cite{Seymour2001ETBoo}, giving a period of about 113~yr and an angular
semimajor axis of 261~mas. Variation of the orbital period is hardly detectable with available
data; therefore, new observations of minima and also astrometry are needed.

\subsection{i Boo}  
\noindent i~Boo (\objectname{HD 133640}, HR~5618, HIP~73695, STF~1909AB) is a well-known EB of the
W~UMa type and, at a distance about 13~pc, also the nearest system in the catalogue. Many times of
minima have been observed over the last 90 years, including three new values listed in Table
\ref{Minima}, but a satisfactory explanation of the period changes is still lacking. It is an X-ray
binary and has also been found to exhibit flares. There is quite a large set of astrometric
measurements of the visual binary, dating back to its discovery by William Herschel in 1781, and
most phases of the orbit are quite well defined. The most recent orbital analysis  finds a period
of about 206~yr and a semimajor axis of 3.8\arcs (\citealt{Soderhjelm1999V640Cas}).

\subsection{V1055 Sco}  
\noindent \objectname{V1055 Sco} (HD~148121, HIP~80603, B~872AB) is a $\beta$~Lyrae EB. There have
been only seven times of minima published in the literature, two of them derived from photometric
data obtained by the Optical Monitoring Camera (OMC) onboard the INTEGRAL satellite (see Table
\ref{Minima}). Astrometric measurements obtained during the last 70 years cover only about
15$^\circ$ in position angle, making it much too premature to attempt a solution to the visual
orbit.

\subsection{V819 Her} 
\noindent \objectname{V819 Her} (HD~157482, HR~6469, HIP~84949, MCA~47) is an Algol-type EB,
orbiting about a common center of mass with a third component in a 5.5-yr period orbit.
Eccentricity is about 0.67, and LITE is evident. The wider pair was discovered by speckle
interferometry in 1980 (\citealt{McAlister1983}) and has been extensively observed by this
technique and also more recently with the Palomar Testbed Interferometer
(\citealt{Muterspaugh2008}). In this system LITE was analyzed together with the interferometry and
radial velocity data (see \citealt{Muterspaugh2006}).

\subsection{V906 Sco}  
\begin{figure}
  \plotone{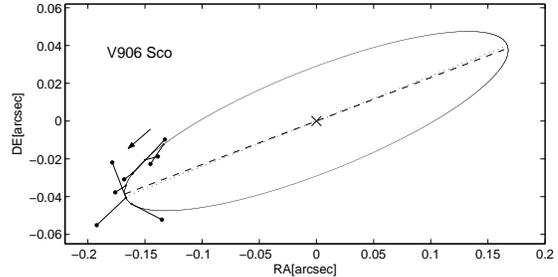}
  \caption{Relative orbit of V906~Sco on the plane of the sky.} \label{V906Sco}
\end{figure}

\noindent \objectname{V906 Sco} (HD~162724, HR~6662, HIP~87616, B~1871AB) is a detached
triple-lined eclipsing binary. A detailed photometric and spectroscopic analysis of this system was
made by \cite{V906Sco1997}, who also included a discussion about possible apsidal motion. A new
visual orbit has been derived (see Fig.\ \ref{V906Sco} and Table \ref{Orbits}); in this solution
only the more precise interferometric measurements from the astrometric data set were used, due to
the much larger scatter in the earlier micrometry data. The last astrometric observations of any
type were obtained more than 15 years ago, so new measurements are needed to improve upon this
solution. Also a total mass about 100\Mo~indicates this orbit to be a preliminary one.

\subsection{V2388 Oph} 
\noindent \objectname{V2388 Oph} (HD~163151, HR~6676, HIP~87655, FIN~381) is a contact eclipsing
binary of W~UMa type. A detailed analysis of $ubvy$ light curves was performed by \cite{Yakut2004}
and a RV analysis by \cite{Rucinski2002}. A preliminary solution for the visual orbit was made by
\cite{Hartkopf1996}, finding a period of about 9~yr. Unfortunately there have been only a few
observations published of times of minima; due to the poor sampling afforded by these data, no
variation in the $O-C$ diagram for minima times is evident.

\subsection{V1647 Sgr}  
\begin{figure}
  \plotone{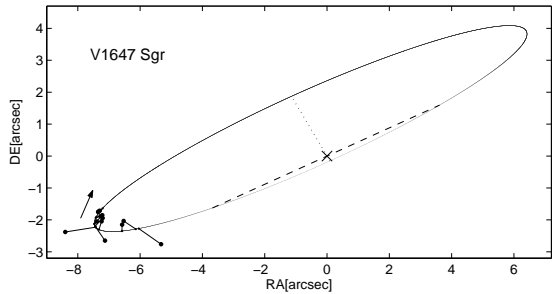}
  \caption{Relative orbit of V1647~Sgr on the plane of the sky.} \label{V1647Sgr}
\end{figure}

\noindent \objectname{V1647 Sgr} (HD~163708, HIP~88069, HJ~5000) is an Algol-type EB. A few dozen
times of minima are available; these data indicate a slow apsidal motion with a period about
530~yr, (see \citealt{Wolf2000V1647Sgr}). \cite{Andersen1985} published the most detailed study of
the whole system to date, including into their analysis also the visual component. Astrometric
measurements of the visual pair obtained over the past 170 years cover only about 14 degrees in
position angle, as shown in Fig.\ \ref{V1647Sgr}. A preliminary orbit was computed for the first
time, but the results are not very convincing due to the limited phase coverage. The period derived
here is about 1200~yr (see Table \ref{Orbits} for the orbital parameters); the total mass which
results in more than 3000\Mo, which makes this solution unrealistic.

\subsection{V772 Her} 
\noindent \objectname{V772 Her} (HD~165590, HIP~88637, STT~341AB) is an eclipsing binary as well as
an RS CVn variable star. This is the A component of a close visual triple and spectroscopic binary
of very high eccentricity; \cite{Heintz1982} derived values of $p_3 = 20$~years and $e=0.956$. See
also \cite{Batten1979} for details on the spectroscopic orbit. The C component is physically
connected to the system and was also found to be a spectroscopic variable (period is about 25 days,
according to \citealt{Fekel1994}). We therefore deal with at least a quintuple system (the WDS
lists four additional wide companions, but these are probably optical due to their fast relative
motion, see \citealt{Tokovinin1997MSC}). There have been more than 20 times of minima published to
date. \cite{Bruton1989} published an $O-C$ diagram with 16 minima times, fitted with a LITE curve
based on parameters derived from the visual orbit.

\subsection{V2083 Cyg}
\noindent \objectname{V2083 Cyg} (HD~184242, HIP~96011, A~713) is an Algol-type EB. There has been
only one time of minimum (based on Hipparcos data) published, our 6 new observations are in Table
\ref{Minima}. Astrometry of the close visual pair during the last century covers about 70$^\circ$
of the orbit. A preliminary orbital solution by \cite{Seymour2002V2083Cyg} gives a period of about
372~yr and an angular semimajor axis about 498~mas.

\subsection{QS Aql} 
\noindent \objectname{QS Aql} (HD~185936, HR~7486, HIP~96840, KUI~93) is an Algol-type eclipsing,
and also spectroscopic, binary. The 24 available minima observations (four of which are new; see
Table \ref{Minima}), allow the period variation to be clearly visible.  The close visual binary has
been observed for over 70 years; the recent orbit by \cite{Docobo2007LOHya} finds an extremely
large eccentricity (e=0.966), but many of the observations show large residuals to this 62-yr
period solution. Combined analysis is still problematic (see \citealt{Mayer2004}) due to poor
coverage of the system by both methods.

\subsection{V505 Sgr}
\noindent \objectname{V505 Sgr} (HD~187949, HR~7571, HIP~97849, CHR~90) is an Algol-type eclipsing
(and also spectroscopic) binary. Since its discovery as a variable, several light-curve
measurements and analyses have been attempted, the latest by \cite{Iban2000}. A detailed spectral
analysis of the system was published by \cite{Tomkin1992}, who also noted discovery of a third
component in the spectrum of the system, as well as a slow change in the radial velocity of this
component. Solution of the visual orbit is still uncertain: \cite{Tomkin1992} estimated a period of
about 100~yr, while \cite{Mayer1997} derived an orbit with a period of about 38~yr; the recent
orbit by \cite{Cvetkovic2008} found a period of 60~yr. The problems of the visual orbit and the
LITE solution were discussed in \cite{Zasche2008}. The set of times of minima is quite large (more
than 300 observations), but a detailed explanation of the period changes is still lacking (see
\citealt{V505Sgr2006}). Nine new minima observations were obtained, see Table \ref{Minima}.

\subsection{MR Del}  
\begin{figure}
  \plotone{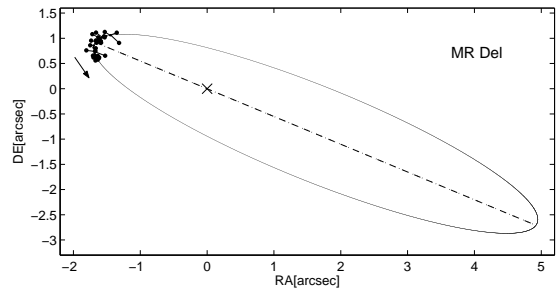}
  \caption{Relative orbit of MR~Del on the plane of the sky.} \label{MRDel}
\end{figure}

\noindent \objectname{MR Del} (HD~195434, HIP~101236, AG~257AB) is an Algol-type EB; at $V = 11.01$
it is the faintest object in the catalog. There have only been a few times of minima observed
during the last 15 years; four additional new measurements are given in Table \ref{Minima}.
Although the visual pair has been observed for a century, the position angle has changed by only
about 15$^\circ$. Such a small change suggests an orbital period of order 2000~yr; our preliminary
solution is given in Table \ref{Orbits} and illustrated in Fig.\ \ref{MRDel}. The resulting total
mass results in about 1.2\Mo.

\subsection{VW Cep}  
\noindent \objectname{VW Cep} (HD~197433, HIP~101750, HEI~7) is a W~UMa-type system, whose primary
and secondary are both chromosphericaly active. There have been numerous light-time effect studies
made of this system; \cite{HerSch} proposed the presence of a third body with an orbital period of
29 years and an angular separation between 0.5\arcs and 1.2\arcs. In 1974, the first successful
visual observation of the third component was obtained by \cite{Heintz1975}. The visual orbit is
fairly well defined, and the parameters of this orbit (see \citealt{Docobo2005}) are in agreement
with the LITE variation in the $O-C$ diagram of minima timings (see ZW for details). This system
seems to be perhaps the most suitable one for simultaneous analysis of both the period variation
and the visual orbit; this technique could also derive the distance to the system. Six new minima
observations were obtained for this paper; see Table \ref{Minima}.

\subsection{BR Ind}
\noindent \objectname{BR Ind} (HD~201427, HIP~104604, HU~1626AB) is an Algol-type EB. The position
angle of the visual binary changed by $\sim$80$^{\circ}$ between 1914 and 2001; this was sufficient
to define a preliminary orbit of $p_3 = 167$~yr and $a = 894$~mas (\citealt{Seymour2002V2083Cyg}).
The only published time of minimum was derived from Hipparcos data.

\subsection{Other systems} \label{Other}
There are also numerous other cases where the strict conditions introduced in the beginning of
Section \ref{Catal} are not satisfied. Some of the more interesting systems are presented in Table
\ref{TableOther}. Their visual orbits have not yet been derived, due to insufficient phase
coverage. In most of these systems the change of position angle is too slow, or there are still
only a few measurements available. One could expect that during the next decades some of these
systems will move to that ones in Chapter \ref{Catal}.

\begin{deluxetable*}{ r l | c c c c l l r r r c c c l }
  \tablewidth{\textwidth}
 \tabletypesize{\scriptsize}
 \tablecaption{Other Eclipsing-Visual Systems. \label{TableOther}}
 \tablehead{ \\[2mm]
  \multicolumn{2}{c|}{\scriptsize Star} & \multicolumn{3}{c}{\scriptsize Spectral types} & \colhead{\scriptsize EB} & \colhead{\scriptsize $V$} & \colhead{\scriptsize $P$} & \colhead{\scriptsize Min} & \colhead{\scriptsize Min} & \colhead{\scriptsize $M$} & \colhead{\scriptsize Depth} & \colhead{\scriptsize Depth} & \colhead{\scriptsize $\Delta \Theta / \Delta T$} &\colhead{\scriptsize References}
  \\[2mm]
  \colhead{\scriptsize HD} & {\tiny Designation$\!\!$} & 1 & 2 & 3 & \colhead{\scriptsize type} & \colhead{\scriptsize [mag]} & \colhead{\scriptsize [day]} & \colhead{\scriptsize Pri} & \colhead{\scriptsize Sec} & \colhead{\scriptsize Astr.} & \colhead{\scriptsize MinP} & \colhead{\scriptsize MinS} & \colhead{\scriptsize $[^\circ] / [\mathrm{yr}]$} & \\[0mm] }
 \\[2mm]
  {\tiny 1061 } &{\tiny   UU Psc     }&  & {\tiny F0IV} &  &{\tiny  EW  }&{\tiny  \phn6.01  }&{\tiny \phn\phn\phn0.84168 }&{\tiny  1  }&{\tiny 0 }&{\tiny 145 }&{\tiny 0.05 V }&{\tiny  0.05 V }&{\tiny \phn1/225 }&{\tiny 1,2,3}\\
  {\tiny 4161 } &{\tiny   YZ Cas     }&{\tiny   A2IV  }&{\tiny   F2V   }&{\tiny       }&{\tiny  EA  }&{\tiny  \phn5.65  }&{\tiny \phn\phn\phn4.46722 }&{\tiny  32 }&{\tiny 3 }&{\tiny  11 }&{\tiny 0.41 B }&{\tiny  0.07 B }&{\tiny \phn0/103 }&{\tiny 4,5}\\
  {\tiny 4502 } &{\tiny $\zeta$ And  }&{\tiny  K1III  }&{\tiny    F    }&{\tiny       }&{\tiny  EW  }&{\tiny  \phn4.10  }&{\tiny \phn\phn   17.7695  }&{\tiny   3 }&{\tiny 2 }&{\tiny  4  }&{\tiny 0.22 V }&{\tiny  0.10 V }&{\tiny 40/91\phn }&{\tiny 4,6}\\
  {\tiny 5679 } &{\tiny   U Cep      }&{\tiny   B7V }&{\tiny G8III-IV }&{\tiny G8III  }&{\tiny  EA  }&{\tiny  \phn6.92  }&{\tiny \phn\phn\phn2.49291 }&{\tiny 1268}&{\tiny 1 }&{\tiny  8  }&{\tiny 2.49 V }&{\tiny  0.18 V }&{\tiny \phn1/119 }&{\tiny 4,7}\\
  {\tiny 8152 } &{\tiny  AQ Psc      }& &{\tiny  F8V  } & &{\tiny  EW  }&{\tiny  \phn8.67  }&{\tiny \phn\phn\phn0.47561 }&{\tiny  11 }&{\tiny11 }&{\tiny  4  }&{\tiny 0.36 V }&{\tiny  0.36 V }&{\tiny  6/81     }&{\tiny 1,8,9}\\
  {\tiny      } &{\tiny  RS Tri      }& &{\tiny   A5V} & &{\tiny  EA  }&{\tiny     10.26  }&{\tiny \phn\phn\phn1.90892 }&{\tiny  85 }&{\tiny 1 }&{\tiny  9  }&{\tiny 0.73 V }&{\tiny  0.13 V }&{\tiny \phn1/110 }&{\tiny 4,10}\\
  {\tiny 13078} &{\tiny  BX And      }&{\tiny   F2V   }&{\tiny    K    }&{\tiny       }&{\tiny  EB  }&{\tiny  \phn8.98  }&{\tiny \phn\phn\phn0.61011 }&{\tiny 257 }&{\tiny42 }&{\tiny 22  }&{\tiny 0.67 b }&{\tiny  0.25 b }&{\tiny \phn2/176 }&{\tiny 4,11}\\
  {\tiny      } &{\tiny  GZ And      }&{\tiny   G5V   }&{\tiny   G5V   }&{\tiny  M3V  }&{\tiny  EW  }&{\tiny     10.89  }&{\tiny \phn\phn\phn0.30502 }&{\tiny  99 }&{\tiny118}&{\tiny  9  }&{\tiny 0.78 V }&{\tiny  0.75 V }&{\tiny \phn4/170 }&{\tiny 4,12,13}\\
  {\tiny      } &{\tiny  EX Per      }&  \multicolumn{2}{c}{\tiny{A1}} &{\tiny   }&{\tiny  EA  }&{\tiny     12.00  }&{\tiny \phn\phn\phn8.47595 }&{\tiny   8 }&{\tiny 2 }&{\tiny 10  }&{\tiny 0.80 V }&{\tiny  0.50 V }&{\tiny  2/96     }&{\tiny 4}\\
  {\tiny 18541} &{\tiny  ST Per      }&{\tiny   A3    }&{\tiny K1-2IV  }&{\tiny       }&{\tiny  EA  }&{\tiny  \phn9.61  }&{\tiny \phn\phn\phn2.64836 }&{\tiny 341 }&{\tiny 2 }&{\tiny  4  }&{\tiny 1.88 V }&{\tiny  0.10 V }&{\tiny  0/70     }&{\tiny 4,14}\\
  {\tiny 24909} &{\tiny  IQ Per      }&{\tiny   B8V   }&{\tiny   A6V   }&{\tiny  A3   }&{\tiny  EA  }&{\tiny  \phn7.73  }&{\tiny \phn\phn\phn1.74356 }&{\tiny 125 }&{\tiny15 }&{\tiny 18  }&{\tiny 0.55 V }&{\tiny  0.16 V }&{\tiny \phn0/132 }&{\tiny 4,15,16}\\
  {\tiny 25638} &{\tiny  SZ Cam      }&{\tiny  O9IV   }&{\tiny  B0.5V  }&{\tiny  B0   }&{\tiny  EB  }&{\tiny  \phn6.93  }&{\tiny \phn\phn\phn2.69847 }&{\tiny  77 }&{\tiny14 }&{\tiny 78  }&{\tiny 0.29 B }&{\tiny  0.24 B }&{\tiny \phn2/177 }&{\tiny 4,17}\\
  {\tiny 34335} &{\tiny  CD Tau      }&{\tiny   F7V   }&{\tiny  F5IV   }&{\tiny  K2   }&{\tiny  EA  }&{\tiny  \phn6.77  }&{\tiny \phn\phn\phn3.43514 }&{\tiny  51 }&{\tiny40 }&{\tiny 29  }&{\tiny 0.57 V }&{\tiny  0.54 V }&{\tiny \phn1/178 }&{\tiny 4,18}\\
  {\tiny 35921} &{\tiny  LY Aur      }&{\tiny  O9III  }&{\tiny O9.5III }&{\tiny  B0V  }&{\tiny  EB  }&{\tiny  \phn6.85  }&{\tiny \phn\phn\phn4.00250 }&{\tiny  32 }&{\tiny17 }&{\tiny 31  }&{\tiny 0.69 V }&{\tiny  0.60 V }&{\tiny \phn4/105 }&{\tiny 4,19}\\
  {\tiny 37020} &{\tiny$\theta$ Ori  }&{\tiny   B3V   }&{\tiny  A7IV   }&{\tiny  O7V  }&{\tiny  EA  }&{\tiny  \phn7.96  }&{\tiny \phn\phn\phn6.47053 }&{\tiny  19 }&{\tiny 0 }&{\tiny 112 }&{\tiny 0.75 V }&{\tiny  0.08 V }&{\tiny \phn0/176 }&{\tiny 4,20}\\
  {\tiny      } &{\tiny              }&{\tiny         }&{\tiny       }&{\tiny }&{\tiny+ EA\phn\phn}&{\tiny + 6.73\phn\phn\phn}&{\tiny $\!\!\!$ + 65.4329\phn }&{\tiny + 24 }&{\tiny +0 }&{\tiny     }&{\tiny 0.93 V }&{\tiny         }&{\tiny           }&{\tiny 4,20}\\
  {\tiny 62863} &{\tiny  PV Pup      }&{\tiny   A8V   }&{\tiny   A8V   }&{\tiny  A2V  }&{\tiny  EA  }&{\tiny  \phn6.93  }&{\tiny \phn\phn\phn1.66073 }&{\tiny   2 }&{\tiny 3 }&{\tiny 59  }&{\tiny 0.44 V }&{\tiny  0.43 V }&{\tiny \phn1/225 }&{\tiny 4,21}\\
  {\tiny 65818} &{\tiny   V Pup      }&{\tiny   B1V   }&{\tiny  B3IV   }&{\tiny       }&{\tiny  EB  }&{\tiny  \phn4.45  }&{\tiny \phn\phn\phn1.45449 }&{\tiny  16 }&{\tiny 4 }&{\tiny  6  }&{\tiny 0.57 V }&{\tiny  0.47 V }&{\tiny \phn1/119 }&{\tiny 4,22}\\
  {\tiny 74307} &{\tiny   S Cnc      }&{\tiny   B9V   }&{\tiny  G8IV   }&{\tiny  G0V  }&{\tiny  EA  }&{\tiny  \phn8.35  }&{\tiny \phn\phn\phn9.48454 }&{\tiny 234 }&{\tiny 0 }&{\tiny  5  }&{\tiny 1.96 V }&{\tiny  0.10 V }&{\tiny \phn0/120 }&{\tiny 4,16,23}\\
  {\tiny 75821} &{\tiny  KX Vel      }& & {\tiny B0III} & &{\tiny  EA  }&{\tiny  \phn5.09  }&{\tiny \phn\phn   26.30624 }&{\tiny  0  }&{\tiny 0 }&{\tiny  8  }&{\tiny 0.08 B }&{\tiny         }&{\tiny  4/65     }&{\tiny 4,24}\\
  {\tiny 83950} &{\tiny   W UMa      }&{\tiny   F8V   }&{\tiny   F8V   }&{\tiny       }&{\tiny  EA  }&{\tiny  \phn7.96  }&{\tiny \phn\phn\phn0.33363 }&{\tiny 1176}&{\tiny284}&{\tiny  4  }&{\tiny 0.73 V }&{\tiny  0.68 V }&{\tiny  7/80     }&{\tiny 4,25}\\
  {\tiny 89714} &{\tiny  HP Car      }& &{\tiny B0.5III } & &{\tiny  EA  }&{\tiny  \phn8.93  }&{\tiny \phn\phn\phn1.60045 }&{\tiny  1  }&{\tiny 0 }&{\tiny  4  }&{\tiny 0.45 V }&{\tiny  0.45 V }&{\tiny 18/59\phn }&{\tiny 4,26}\\
  {\tiny 93206} &{\tiny  QZ Car      }&{\tiny  O9.7I  }&{\tiny   B0I   }&{\tiny  O9V  }&{\tiny  EB  }&{\tiny  \phn6.24  }&{\tiny \phn\phn\phn5.9991  }&{\tiny  5  }&{\tiny 1 }&{\tiny  4  }&{\tiny 0.33 V }&{\tiny  0.27 V }&{\tiny  0/85     }&{\tiny 4,27,28}\\
  {\tiny      } &{\tiny  AM Leo      }& & {\tiny F8V } & &{\tiny  EW  }&{\tiny  \phn9.31  }&{\tiny \phn\phn\phn0.36580 }&{\tiny 194 }&{\tiny139}&{\tiny 41  }&{\tiny 0.58 V }&{\tiny  0.58 V }&{\tiny \phn0/175 }&{\tiny 4,29}\\
  {\tiny 99769} &{\tiny  MN Cen      }& & {\tiny B2-3V} & &{\tiny  EA  }&{\tiny  \phn8.8   }&{\tiny \phn\phn\phn3.48901 }&{\tiny  7  }&{\tiny 4 }&{\tiny 10  }&{\tiny 0.40 b }&{\tiny  0.10 b }&{\tiny \phn0/117 }&{\tiny 4,30}\\
  {\tiny 99946} &{\tiny  AW UMa      }& \multicolumn{2}{c}{\tiny{A8Vn}} &{\tiny G5}&{\tiny  EW  }&{\tiny  \phn6.92  }&{\tiny \phn\phn\phn0.43873 }&{\tiny 103 }&{\tiny72 }&{\tiny 21  }&{\tiny 0.30 V }&{\tiny  0.25 V }&{\tiny \phn0/131 }&{\tiny 4,31}\\
  {\tiny106400} &{\tiny  AH Vir      }&{\tiny   G8V   }&{\tiny   G8V   }&{\tiny  K1V  }&{\tiny  EW  }&{\tiny  \phn9.33  }&{\tiny \phn\phn\phn0.40753 }&{\tiny 261 }&{\tiny169}&{\tiny  7  }&{\tiny 0.60 V }&{\tiny  0.53 V }&{\tiny  2/93     }&{\tiny 4,32}\\
  {\tiny114911} &{\tiny $\eta$ Mus   }& & {\tiny B8V} & &{\tiny  EA  }&{\tiny  \phn4.77  }&{\tiny \phn\phn\phn2.3963  }&{\tiny  2  }&{\tiny 0 }&{\tiny 17  }&{\tiny 0.05 V }&{\tiny         }&{\tiny \phn3/176 }&{\tiny 4,33}\\
  {\tiny117408} &{\tiny  SS Hya      }& \multicolumn{2}{c}{\tiny{A0V}} &{\tiny A0 }&{\tiny  EA  }&{\tiny  \phn7.86  }&{\tiny \phn\phn\phn8.2     }&{\tiny  1  }&{\tiny 0 }&{\tiny  4  }&{\tiny 0.22 B }&{\tiny         }&{\tiny \phn6/108 }&{\tiny 4,34}\\
  {\tiny132742} &{\tiny$\delta$ Lib  }&{\tiny   A0V   }&{\tiny   K0IV  }&{\tiny  F    }&{\tiny  EA  }&{\tiny  \phn4.95  }&{\tiny \phn\phn\phn2.32735 }&{\tiny 200 }&{\tiny 0 }&{\tiny  4  }&{\tiny 0.99 V }&{\tiny         }&{\tiny  3/94     }&{\tiny 4,35}\\
  {\tiny134646} &{\tiny              }& \multicolumn{2}{c}{\tiny{F4III}} &{\tiny G8}&{\tiny EA  }&{\tiny  \phn6.82  }&{\tiny \phn\phn\phn2.44405 }&{\tiny  1  }&{\tiny 0 }&{\tiny 11  }&{\tiny 0.13 V }&{\tiny  0.06 V }&{\tiny \phn2/102 }&{\tiny 36,37}\\
  {\tiny135421} &{\tiny BV Dra       }&{\tiny   F9V   }&{\tiny   F8V   }&{\tiny       }&{\tiny  EW  }&{\tiny  \phn8.04  }&{\tiny \phn\phn\phn0.35007 }&{\tiny  53 }&{\tiny61 }&{\tiny 39  }&{\tiny 0.60 V }&{\tiny         }&{\tiny \phn1/167 }&{\tiny 4,38}\\
  {\tiny      } &{\tiny   + BW Dra   }&{\tiny   + G3V }&{\tiny + G0V  }&{\tiny }&{\tiny+ EW\phn\phn}&{\tiny + 8.74\phn\phn\phn}&{\tiny\phn + 0.29217 }&{\tiny+ 53}&{\tiny+ 42}&{\tiny     }&{\tiny+ 0.47 V}&{\tiny+ 0.41 V }&{\tiny           }&{\tiny 4,38}\\
  {\tiny138672} &{\tiny  EI Lib      }&{\tiny   A3-7  }&{\tiny   F5-8  }&{\tiny  Am   }&{\tiny  EA  }&{\tiny  \phn9.50  }&{\tiny \phn\phn\phn1.98691 }&{\tiny  4  }&{\tiny 0 }&{\tiny  7  }&{\tiny 1.00 b }&{\tiny         }&{\tiny  3/79     }&{\tiny 4,16,34}\\
  {\tiny139319} &{\tiny  TW Dra      }&{\tiny   A5V   }&{\tiny  K0III  }&{\tiny       }&{\tiny  EA  }&{\tiny  \phn7.43  }&{\tiny \phn\phn\phn2.80685 }&{\tiny 499 }&{\tiny 5 }&{\tiny 36  }&{\tiny 2.50 b }&{\tiny         }&{\tiny \phn4/157 }&{\tiny 4,39}\\
  {\tiny139966} &{\tiny  HH Nor      }& \multicolumn{2}{c}{\tiny{F0}} &{\tiny F0IV}&{\tiny  EA  }&{\tiny     10.2   }&{\tiny \phn\phn\phn8.58313 }&{\tiny  1  }&{\tiny 0 }&{\tiny  9  }&{\tiny 1.20 b }&{\tiny         }&{\tiny \phn0/164 }&{\tiny 4,30,40}\\
  {\tiny149730} &{\tiny   R Ara      }& & {\tiny B9IV} & &{\tiny  EA  }&{\tiny  \phn6.65  }&{\tiny \phn\phn\phn4.42509 }&{\tiny  6  }&{\tiny 0 }&{\tiny 21  }&{\tiny 0.90 b }&{\tiny  0.20 b }&{\tiny \phn8/164 }&{\tiny 4,30}\\
  {\tiny150708} &{\tiny  WW Dra      }&{\tiny   G2IV  }&{\tiny   K0IV  }&{\tiny F2V   }&{\tiny  EA  }&{\tiny  \phn8.59  }&{\tiny \phn\phn\phn4.62972 }&{\tiny 84  }&{\tiny10 }&{\tiny 23  }&{\tiny 0.65 V }&{\tiny  0.08 V }&{\tiny \phn1/175 }&{\tiny 4,20}\\
  {\tiny153751} &{\tiny$\epsilon$ UMi}&{\tiny  G1    }&{\tiny   A-F   }&{\tiny       }&{\tiny  EA  }&{\tiny  \phn4.22  }&{\tiny \phn\phn   39.4809  }&{\tiny  1  }&{\tiny 0 }&{\tiny  6  }&{\tiny 0.04 V }&{\tiny  0.02 V }&{\tiny  4/80     }&{\tiny 4,41}\\
  {\tiny155937} &{\tiny  AK Her      }&{\tiny   F2V   }&{\tiny   F6V   }&{\tiny K2V   }&{\tiny  EW  }&{\tiny  \phn8.51  }&{\tiny \phn\phn\phn0.42152 }&{\tiny 460 }&{\tiny161}&{\tiny 14  }&{\tiny 0.48 V }&{\tiny  0.35 V }&{\tiny \phn0/104 }&{\tiny 4,12,20}\\
  {\tiny156247} &{\tiny   U Oph      }&{\tiny   B5V   }&{\tiny   B5V   }&{\tiny  B5   }&{\tiny  EA  }&{\tiny  \phn5.90  }&{\tiny \phn\phn\phn1.67735 }&{\tiny 431 }&{\tiny257}&{\tiny 13  }&{\tiny 0.72 V }&{\tiny  0.62 V }&{\tiny \phn3/123 }&{\tiny 4,42}\\
  {\tiny156633} &{\tiny   u Her      }&{\tiny   B2IV  }&{\tiny  B8III  }&{\tiny  B    }&{\tiny  EB  }&{\tiny  \phn4.80  }&{\tiny \phn\phn\phn2.05103 }&{\tiny 222 }&{\tiny36 }&{\tiny 28  }&{\tiny 0.68 V }&{\tiny  0.24 V }&{\tiny \phn2/158 }&{\tiny 4,43}\\
  {\tiny161321} &{\tiny V624 Her     }&{\tiny   A4    }&{\tiny   A7    }&{\tiny       }&{\tiny  EA  }&{\tiny  \phn6.20  }&{\tiny \phn\phn\phn3.89498 }&{\tiny  3  }&{\tiny 1 }&{\tiny  7  }&{\tiny 0.18 V }&{\tiny  0.17 V }&{\tiny  0/98     }&{\tiny 4,44}\\
  {\tiny161783} &{\tiny V539 Ara     }&{\tiny   B3V   }&{\tiny   B4V   }&{\tiny A0-1V }&{\tiny  EA  }&{\tiny  \phn5.92  }&{\tiny \phn\phn\phn3.16913 }&{\tiny 17  }&{\tiny12 }&{\tiny 16  }&{\tiny 0.52 V }&{\tiny  0.43 V }&{\tiny \phn2/166 }&{\tiny 4,45}\\
  {\tiny163181} &{\tiny V453 Sco     }& & {\tiny B0I} & &{\tiny  EB  }&{\tiny  \phn6.61  }&{\tiny \phn\phn   12.00597 }&{\tiny  4  }&{\tiny 1 }&{\tiny  5  }&{\tiny 0.37 V }&{\tiny  0.34 V }&{\tiny  3/89     }&{\tiny 4,33}\\
  {\tiny165814} &{\tiny V3792 Sgr    }& \multicolumn{2}{c}{\tiny{B4IV}} &{\tiny B9IV}&{\tiny EB  }&{\tiny  \phn6.71  }&{\tiny \phn\phn\phn2.24808 }&{\tiny  2  }&{\tiny 0 }&{\tiny 10  }&{\tiny 0.45 V }&{\tiny  0.38 V }&{\tiny \phn1/130 }&{\tiny 4,34,46}\\
  {\tiny166937} &{\tiny $\mu$ Sgr    }&{\tiny   B8I   }&{\tiny  B1.5V  }&{\tiny B9III }&{\tiny  EA  }&{\tiny  \phn3.84  }&{\tiny \phn      180.55    }&{\tiny  1  }&{\tiny 0 }&{\tiny 18  }&{\tiny 0.08 V }&{\tiny         }&{\tiny \phn5/177 }&{\tiny 1,42,47}\\
  {\tiny      } &{\tiny  TZ Lyr      }& & {\tiny F5V} & &{\tiny  EB  }&{\tiny     10.77  }&{\tiny \phn\phn\phn0.52883 }&{\tiny 400 }&{\tiny 5 }&{\tiny  5  }&{\tiny 0.98 V }&{\tiny  0.18 V }&{\tiny  6/74     }&{\tiny 4,29}\\
  {\tiny167647} &{\tiny  RS Sgr      }&{\tiny   B5V   }&{\tiny   A2V   }&{\tiny A1V   }&{\tiny  EA  }&{\tiny  \phn6.03  }&{\tiny \phn\phn\phn2.41568 }&{\tiny 36  }&{\tiny 8 }&{\tiny  8  }&{\tiny 0.96 V }&{\tiny  0.27 V }&{\tiny \phn2/108 }&{\tiny 4,48,49}\\
  {\tiny174638} &{\tiny $\beta$ Lyr  }&{\tiny   B8II  }&{\tiny  B6.5   }&{\tiny B5V   }&{\tiny  EB  }&{\tiny  \phn3.52  }&{\tiny \phn\phn   12.93763 }&{\tiny 1475}&{\tiny923}&{\tiny 93  }&{\tiny 1.11 V }&{\tiny  0.60 V }&{\tiny \phn6/129 }&{\tiny 4,50,51}\\
  {\tiny178001} &{\tiny  BH Dra      }&{\tiny   A2V   }&{\tiny    A7   }&{\tiny Am    }&{\tiny  EA  }&{\tiny  \phn8.43  }&{\tiny \phn\phn\phn1.81724 }&{\tiny 193 }&{\tiny 0 }&{\tiny 20  }&{\tiny 0.89 V }&{\tiny  0.20 V }&{\tiny \phn2/103 }&{\tiny 4,10,33}\\
  {\tiny180639} &{\tiny V342 Aql     }&{\tiny  A4II   }&{\tiny  K0IV   }&{\tiny K2II  }&{\tiny  EA  }&{\tiny  \phn8.68  }&{\tiny \phn\phn\phn3.39088 }&{\tiny 121 }&{\tiny 0 }&{\tiny 41  }&{\tiny 3.40 b }&{\tiny         }&{\tiny \phn1/160 }&{\tiny 4,52,53}\\
  {\tiny181987} &{\tiny   Z Vul      }&{\tiny   B4V   }&{\tiny  A3III  }&{\tiny       }&{\tiny  EA  }&{\tiny  \phn7.33  }&{\tiny \phn\phn\phn2.45493 }&{\tiny 425 }&{\tiny17 }&{\tiny  5  }&{\tiny 1.65 V }&{\tiny  0.33 V }&{\tiny \phn0/101 }&{\tiny 4,54}\\
  {\tiny183794} &{\tiny  V822~Aql    }&{\tiny   B3    }&{\tiny  B9     }&{\tiny       }&{\tiny  EA  }&{\tiny  \phn7.12  }&{\tiny \phn\phn\phn5.29495 }&{\tiny  16 }&{\tiny 6 }&{\tiny 24  }&{\tiny 0.57 V }&{\tiny  0.20 V }&{\tiny \phn5/126 }&{\tiny 4,73}\\
  {\tiny185507} &{\tiny$\sigma$ Aql  }&{\tiny   B3V   }&{\tiny  B3V    }&{\tiny G8V   }&{\tiny  EB  }&{\tiny  \phn5.18  }&{\tiny \phn\phn\phn1.95026 }&{\tiny  6  }&{\tiny 3 }&{\tiny  6  }&{\tiny 0.20 V }&{\tiny  0.10 V }&{\tiny \phn3/170 }&{\tiny 4,16,42}\\
  {\tiny191515} &{\tiny V346 Aql     }&{\tiny   A0V   }&{\tiny   G4IV  }&{\tiny       }&{\tiny  EA  }&{\tiny  \phn9.08  }&{\tiny \phn\phn\phn1.10636 }&{\tiny 702 }&{\tiny 0 }&{\tiny 19  }&{\tiny 1.10 b }&{\tiny  0.10 b }&{\tiny  3/69     }&{\tiny 4,10}\\
  {\tiny192909} &{\tiny V1488 Cyg    }&{\tiny   K3I   }&{\tiny   B4-5  }&{\tiny  A    }&{\tiny  EA  }&{\tiny  \phn4.02  }&{\tiny  $\!\!\!\!\!\!$        1147.4     }&{\tiny  1  }&{\tiny 0 }&{\tiny 10  }&{\tiny 0.24 V }&{\tiny         }&{\tiny \phn2/178 }&{\tiny 4,55}\\
  {\tiny      } &{\tiny V1191 Cyg    }& & {\tiny F6V} & &{\tiny  EW  }&{\tiny     10.8   }&{\tiny \phn\phn\phn0.31338 }&{\tiny 23  }&{\tiny19 }&{\tiny  5  }&{\tiny 0.33 V }&{\tiny  0.29 V }&{\tiny \phn2/107 }&{\tiny 4,56}\\
  {\tiny198287} &{\tiny V367 Cyg     }&{\tiny   A7    }&{\tiny  B6-7   }&{\tiny       }&{\tiny  EB  }&{\tiny  \phn7.04  }&{\tiny \phn\phn   18.59773 }&{\tiny 88  }&{\tiny32 }&{\tiny  7  }&{\tiny 0.93 V }&{\tiny  0.49 V }&{\tiny115/56\phn\phn}&{\tiny 4,57}\\
  {\tiny199005} &{\tiny  KZ Pav      }&{\tiny   F6V   }&{\tiny  K4IV   }&{\tiny  F2   }&{\tiny  EA  }&{\tiny  \phn7.75  }&{\tiny \phn\phn\phn0.94988 }&{\tiny 86  }&{\tiny11 }&{\tiny 21  }&{\tiny 1.59 V }&{\tiny         }&{\tiny \phn1/165 }&{\tiny 4,58}\\
  {\tiny203069} &{\tiny  RY Aqr      }&{\tiny   A8V   }&{\tiny   K1    }&{\tiny K2V   }&{\tiny  EA  }&{\tiny  \phn8.86  }&{\tiny \phn\phn\phn1.96658 }&{\tiny 161 }&{\tiny 4 }&{\tiny  4  }&{\tiny 1.30 V }&{\tiny         }&{\tiny \phn1/101 }&{\tiny 4,59}\\
  {\tiny204215} &{\tiny  KP Peg      }& & {\tiny A2V} & &{\tiny  EB  }&{\tiny  \phn7.28  }&{\tiny \phn\phn\phn0.72721 }&{\tiny  8  }&{\tiny 0 }&{\tiny 91  }&{\tiny 0.21 V }&{\tiny         }&{\tiny \phn5/177 }&{\tiny 4,60}\\
  {\tiny206155} &{\tiny  EE Peg      }&{\tiny   A3V   }&{\tiny   F5V   }&{\tiny K5V   }&{\tiny  EA  }&{\tiny  \phn6.98  }&{\tiny \phn\phn\phn2.62821 }&{\tiny 77  }&{\tiny 2 }&{\tiny  5  }&{\tiny 0.58 V }&{\tiny  0.13 V }&{\tiny123/24\phn\phn}&{\tiny 4,20}\\
  {\tiny208392} &{\tiny  EM Cep      }&{\tiny  B0.5V  }&{\tiny   B1V   }&{\tiny B1V   }&{\tiny  EW  }&{\tiny  \phn7.03  }&{\tiny \phn\phn\phn0.80619 }&{\tiny 27  }&{\tiny24 }&{\tiny 26  }&{\tiny 0.15 V }&{\tiny  0.14 V }&{\tiny \phn0/175 }&{\tiny 4,10,61}\\
  {\tiny209278} &{\tiny  DX Aqr      }&{\tiny   A2V   }&{\tiny  K0III  }&{\tiny       }&{\tiny  EA  }&{\tiny  \phn6.37  }&{\tiny \phn\phn\phn0.94501 }&{\tiny 164 }&{\tiny 5 }&{\tiny 81  }&{\tiny 0.41 V }&{\tiny         }&{\tiny \phn2/184 }&{\tiny 4,62}\\
  {\tiny209943} &{\tiny V376 Cep     }& \multicolumn{2}{c}{\tiny{F5}} &{\tiny F6IV-V}&{\tiny EA  }&{\tiny  \phn7.43  }&{\tiny \phn\phn\phn1.166   }&{\tiny  2  }&{\tiny 0 }&{\tiny 150 }&{\tiny 0.07 V }&{\tiny         }&{\tiny \phn9/182 }&{\tiny 1,63}\\
  {\tiny211853} &{\tiny  GP Cep      }&{\tiny WN6o/WCE }&{\tiny  O3-6   }&{\tiny B0:I  }&{\tiny  EB  }&{\tiny  \phn9.03  }&{\tiny \phn\phn\phn6.6884  }&{\tiny  1  }&{\tiny 0 }&{\tiny 12  }&{\tiny 0.11 V }&{\tiny         }&{\tiny  0/94     }&{\tiny 4,64}\\
  {\tiny      } &{\tiny              }&{\tiny         }&{\tiny    }&{\tiny + B1:V-III }&{\tiny + EB\phn\phn  }&{\tiny   }&{\tiny \phn      + 3.4696  }&{\tiny + 1 }&{\tiny+ 0}&{\tiny     }&{\tiny        }&{\tiny         }&{\tiny           }&{\tiny }\\
  {\tiny215661} &{\tiny  ZZ Cep      }&{\tiny   B7    }&{\tiny   F0V   }&{\tiny  A2   }&{\tiny  EA  }&{\tiny  \phn9.00  }&{\tiny \phn\phn\phn2.14180 }&{\tiny 240 }&{\tiny 0 }&{\tiny 23  }&{\tiny 0.95 V }&{\tiny  0.14 V }&{\tiny \phn1/155 }&{\tiny 4,10,65}\\
  {\tiny216309} &{\tiny  SU Aqr      }& & {\tiny A2IV} & &{\tiny  EB  }&{\tiny  \phn9.95  }&{\tiny \phn\phn\phn1.04470 }&{\tiny 44  }&{\tiny 0 }&{\tiny  9  }&{\tiny 0.60 b }&{\tiny  0.30 b }&{\tiny  1/99     }&{\tiny 4,34}\\
  {\tiny219113} &{\tiny  SZ Psc      }&{\tiny  F8V-IV }&{\tiny  K1IV   }&{\tiny       }&{\tiny  EA  }&{\tiny  \phn7.44  }&{\tiny \phn\phn\phn3.96579 }&{\tiny 60  }&{\tiny 0 }&{\tiny  5  }&{\tiny 0.54 V }&{\tiny  0.20 V }&{\tiny  1/91     }&{\tiny 4,66,67}\\
  {\tiny221253} &{\tiny  AR Cas      }& \multicolumn{2}{c}{\tiny{B4.2IV}} &{\tiny A6V}&{\tiny EA }&{\tiny \phn4.88   }&{\tiny \phn\phn\phn6.06633 }&{\tiny  9  }&{\tiny 1 }&{\tiny 11  }&{\tiny 0.14 V }&{\tiny  0.04 V }&{\tiny \phn2/119 }&{\tiny 4,68}\\
  {\tiny232121} &{\tiny  SX Cas      }&{\tiny   B7    }&{\tiny  K3III  }&{\tiny       }&{\tiny  EA  }&{\tiny  \phn9.05  }&{\tiny \phn\phn   36.56375 }&{\tiny  82 }&{\tiny 1 }&{\tiny  6  }&{\tiny 0.87 V }&{\tiny  0.36 V }&{\tiny  1/95     }&{\tiny 4,69}\\
  {\tiny256320} &{\tiny  FI Ori      }&{\tiny   F5    }&{\tiny  K2IV   }&{\tiny F7V   }&{\tiny  EA  }&{\tiny     10.3   }&{\tiny \phn\phn\phn4.44815 }&{\tiny  40 }&{\tiny 1 }&{\tiny 10  }&{\tiny 0.90 b }&{\tiny         }&{\tiny  7/95     }&{\tiny 4,70,71}\\
  {\tiny261025} &{\tiny  AK Aur      }&{\tiny   A1    }&{\tiny  F5III  }&{\tiny  F5   }&{\tiny  EA  }&{\tiny     10.20  }&{\tiny \phn\phn\phn4.76314 }&{\tiny  8  }&{\tiny 1 }&{\tiny 10  }&{\tiny 0.50 B }&{\tiny         }&{\tiny \phn0/179 }&{\tiny 4,10}\\
  {\tiny349425} &{\tiny  AD Her      }&{\tiny   A4V   }&{\tiny   K2    }&{\tiny       }&{\tiny  EA  }&{\tiny  \phn9.68  }&{\tiny \phn\phn\phn9.76661 }&{\tiny 26  }&{\tiny 0 }&{\tiny 17  }&{\tiny 1.47 V }&{\tiny  0.09 V }&{\tiny \phn0/107 }&{\tiny 4,72 }\\[1mm]
\tablecomments{For the explanation of the individual columns, see Table \ref{Table1}. In the column
the 'EB~type' are the following abbreviations: 'EW' for W~UMa type, 'EB' for $\beta$~Lyrae type,
and 'EA' for Algol type eclipsing binaries, respectively. The column $\Delta \Theta / \Delta T$
denotes the mean rate of the position angle variation with time, which has been derived only from
the first and the last measurements of the system. $\theta$~Ori, BV+BW~Dra, and GP~Cep constitute
multiple systems with two eclipsing binaries. The system GP~Cep is more complicated, its light
curve is rather 'eclipsing-like' (see \citealt{2002Demers}). References: (1) \cite{GCVS2004}; (2)
\cite{Cowley1974PASP}; (3) \cite{Shobbrook2005JAD}; (4) \cite{Malkov2006}; (5) \cite{Celikel1989};
(6) \cite{Kovari2007}; (7) \cite{Plavec1983}; (8) \cite{Djurasevic2006}; (9) \cite{Sarma1982}; (10)
\cite{Brancewicz1980}; (11) \cite{Samec1989}; (12) \cite{Rucinski2007}; (13)
\cite{Tokovinin1997MSC}; (14) \cite{Olson1982}; (15) \cite{Lacy1985}; (16) \cite{Halbedel1985};
(17) \cite{Lorenz1998}; (18) \cite{Ribas1999}; (19) \cite{Drechsel1989}; (20) \cite{Chambliss1992};
(21) \cite{Vaz1984}; (22) \cite{Frieboes1962}; (23) \cite{Etzel1985}; (24) \cite{Morgan1955}; (25)
\cite{Rucinski1993}; (26) \cite{Proust1981}; (27) \cite{Walborn1973II}; (28) \cite{Leung1979}; (29)
\cite{Hill1975}; (30) \cite{Houk1975}; (31) \cite{Abt2004}; (32) \cite{Lu1993}; (33)
\cite{Levato1975}; (34) \cite{Houk1988}; (35) \cite{Bakis2006}; (36) \cite{Pagel1960}; (37)
\cite{2007OEJV72}; (38) \cite{Batten1989}; (39) \cite{Zejda2008}; (40) \cite{Budding1984}; (41)
\cite{Hinderer1957}; (42) \cite{Lindroos1985}; (43) \cite{Hilditch2005}; (44) \cite{Petrie1950};
(45) \cite{Andersen83}; (46) \cite{Gray2006AJ}; (47) \cite{Polidan1984}; (48) \cite{Ferrer1986};
(49) \cite{Lindroos1986}; (50) \cite{Guetter1968}; (51) \cite{Harmanec1993}; (52)
\cite{Stephenson1960}; (53) \cite{Erdem2007NewA}; (54) \cite{Popper1957}; (55) \cite{Ginestet2002};
(56) \cite{Rucinski2008}; (57) \cite{Pavlovski1992}; (58) \cite{Svechnikov1990}; (59)
\cite{Helt1987}; (60) \cite{Pych2004}; (61) \cite{Simonson1968}; (62) \cite{Hoffleit1982}; (63)
\cite{Grosheva2006}; (64) \cite{2002Demers}; (65) \cite{Carrier2002}; (66) \cite{Zhang2008}; (67)
\cite{Eaton2007}; (68) \cite{Krylov2003}; (69) \cite{Plavec1982}; (70) \cite{Fehrenbach1961}; (71)
\cite{Budding2004}; (72) \cite{Batten1978}; (73) \cite{Popper1981};}
\end{deluxetable*}

\subsection{Special cases} \label{Special}

There are also a number of ``special cases" which were not included in the catalog due to their
more complicated or uncertain nature. A few examples are described below:

\begin{itemize}
\item In all systems included in the catalog, the close EB comprises one component of a much wider
visual/interferometric pair. However, in a few rare cases the components of the eclipsing binary
also comprise the components of the visual pair. The systems $\beta$~Aur, $\beta$~Cap,
$\gamma$~Per, and V695 Cyg are eclipsing pairs which have also been resolved by interferometry
(another possible example of such a system is $\alpha$~Com). The chances of an orbit being
sufficently edge-on to produce eclipses are of course greater for systems with smaller separations;
it is therefore expected that the number of such ``visual--eclipsing" binaries will increase as
more EBs are observed by long-baseline interferometers. However, these systems do not meet the
criteria for inclusion in this catalog.

\item Another class of objects not included in the catalog is that of the so-called ellipsoidal
variables. Systems such as HD~178125 (18~Aql, Y~Aql) or HD~22124 (IX~Per) are also sometimes
classified as EBs and a few ''minima'' have been published, as well.

\item The system HD 217675 ($o$~And, 1~And) is sometimes classified as an eclipsing binary. In
fact, this quadruple system \citep{1988Hill} is photometrically variable \citep{1972Olsen}, but,
according to \citep{1997Pavlovski}, it is not an eclipsing binary.
\end{itemize}

\section{Discussions and conclusion}

More than 13000 systems in the WDS had sufficiently large astrometric data sets to allow potential
analysis and were, therefore, checked for the presence of EB components. Many stars which were
suspected to be variables were found in this large set, but only a very limited number of these
systems had ever been given detailed spectroscopic or photometric analysis. Systems chosen for
inclusion in this catalog were selected on the basis of EB designations in the Simbad database and
notes in the WDS\footnote{$\mathrm{http://ad.usno.navy.mil/wds/wdsnewnotes\_main.txt}$}. However,
in many cases identification of a star as an eclipsing variable is a rather difficult task; because
of this, many such systems are designated in Simbad only as ``variable stars''. Although the
current number of known visual doubles containing eclipsing variables as components is still quite
limited, this number is expected to increase substantially as the true nature of more of these
``variable stars'' is determined through further photometric observation.

A long-term goal is to increase the size of this catalog to the point where reliable statistical
analysis of this class of systems may be attempted. The subset of visual multiple systems including
eclipsing binary components could be another area of potential interest. If one has information
about the various orbits in these systems, the ratio of periods or the mutual inclinations of the
long and the short orbits could prove an interesting probe into the mechanisms for formation of
these objects. Also a frequency of quadruple or quintuple systems among multiples could be studied.
The main catalog in the present paper includes 7 quadruple systems, 8 quintuples and 1 sextuple.
However, there are still many cases, where the membership of a star to the system is questionable,
so this fraction of multiples is expected to increase.

Regrettably, many of the systems in the catalog lack recent observations. Ironically, this is due
in part to the fact many of these systems are too bright for modern photometric equipment! Some of
the earliest known eclipsing binaries are now neglected, since they can easily saturate a CCD
detector mounted on even a modest telescope. Phase coverage of most visual binary orbits is
insufficient, due largely to the exceedingly long orbital periods of these pairs. Astrometry of
closer interferometric pairs is also lacking, as shifting priorities of telescope allocation
committes has made it difficult for observers to get time on the large telescopes needed for
obtained these data. As a result, analysis is complicated, and newly derived orbital elements are
affected by relatively large errors due to small arcs of the orbit covered and/or sparse phase
coverage.

In conclusion, although a few of the systems in the catalog presented here (e.g., V505~Sgr, QS~Aql,
V2388~Oph) are suitable for simultaneous analysis of period variation and astrometry, in most cases
the time span of observations is still too short and more data are needed. The highest priority
systems for which interferometric observations are desired include V1031~Ori, ET~Boo, and V906~Sco.
Systems especially in need of additional photometric observations for minima determinations are
V1031~Ori, LO~Hya, V906~Sco, and V2388~Oph.

 \acknowledgments
Some of this work was based on data from the OMC Archive at LAEFF, pre-processed by ISDC. We would
like to thank Dr.Milo\v{s} Zejda for kindly sending us the data. Dr. Andrei Tokovinin and also an
anonymous referee are greatly acknowledged for their helpful and critical suggestions. This
investigation was supported by the Czech Science Foundation, grants No. 205/06/0217 and No.
205/06/0304. We also acknowledge the support from the Research Program MSMT0021620860 of the
Ministry of Education and also from the Mexican grant PAPIIT IN113308. This research has made use
of the SIMBAD database, operated at CDS, Strasbourg, France, and of NASA's Astrophysics Data System
Bibliographic Services.


\begin{thebibliography}{}

\bibitem[Abt(1981)]{Abt1981} Abt, H.~A.\ 1981, \apjs, 45, 437

\bibitem[Abt(1985)]{Abt1985} Abt, H.~A.\ 1985, \apjs, 59, 95

\bibitem[Abt(2004)]{Abt2004} Abt, H.~A.\ 2004, \apjs, 155, 175

\bibitem[{{Alencar} {et~al.}(1997){Alencar}, {Vaz}, \& {Helt}}]{V906Sco1997}
{Alencar}, S.~H.~P., {Vaz}, L.~P.~R., \& {Helt}, B.~E. 1997, \aap, 326, 709

\bibitem[{{Alzner} \& {Argyle}(2000)}]{Alzner2000DelVel}
{Alzner}, A. \& {Argyle}, R. 2000, IAU Comm.\ 26, Inf.\ Circ.\ 142

\bibitem[{{Andersen}(1983)}]{Andersen83} {Andersen}, J. 1983, \aap, 118, 255

\bibitem[{{Andersen et al.}(1984)}]{Andersen1984}
Andersen, J., Clausen, J.~V., \& Nordstrom, B.\ 1984, \aap, 134, 147

\bibitem[{{Andersen \& Gimenez}(1985)}]{Andersen1985} Andersen, J., \& Gimenez, A.\ 1985, \aap, 145, 206

\bibitem[{{Andersen} {et~al.}(1990){Andersen}, {Nordstr\"om}, \&
  {Clausen}}]{Andersen1990V1031Ori}
{Andersen}, J., {Nordstr\"om}, B., \& {Clausen}, J.~V. 1990, \aap, 228, 365

\bibitem[Appenzeller(1967)]{Appenzeller1967} Appenzeller, I.\ 1967, \pasp, 79, 102

\bibitem[Argyle et al.(2002)]{Argyle2002} Argyle, R.~W., Alzner, A., \& Horch, E.~P.\ 2002, \aap, 384, 171

\bibitem[{{Aristidi} {et~al.}(1999){Aristidi}, {Prieur}, {Scardia}, {Koechlin},
{Avila}, {Carbillet}, {Lopez}, {Rabbia}, {Nisenson}, \& {Gezari}}]{Aristidi1999DNUMa} {Aristidi},
{\'E}., {Prieur}, J.-L., {Scardia}, M., {Koechlin}, L., {Avila}, R., {Carbillet}, M., {Lopez}, B.
{Rabbia}, Y., {Nisenson}, P., {Gezari}, D. 1999, \aaps, 134, 545

\bibitem[Bak{\i}{\c s} et al.(2006)]{Bakis2006} Bak{\i}{\c s}, V., Budding, E., Erdem,
A., Bak{\i}{\c s}, H., Demircan, O., \& Hadrava, P.\ 2006, \mnras, 370, 1935

\bibitem[{{Bakos}(1985)}]{Bakos1985LOHya} {Bakos}, G.~A. 1985, \jrasc, 79, 119

\bibitem[{Balega {et~al.}(1999)}]{Balega1999}
Balega, I.I., Balega, Yu.Yu., Hofmann, K.-H., Tokovinin, A.A., \& Weigelt, G.\ 1999, SvAL, 25, 797

\bibitem[Bartkevi{\v c}ius \& Gudas(2002)]{ETBoo2002BaltA} Bartkevi{\v c}ius, A.,
\& Gudas, A.\ 2002, Baltic Astronomy, 11, 153

\bibitem[Batten(1973)]{1973Batten} Batten, A.~H.\ 1973, Binary and
multiple systems of stars / by Alan H.~Batten.~Oxford ; New York : Pergamon Press, [1973]
(International series of monographs in natural philosophy ; v.~51),

\bibitem[Batten \& Fletcher(1978)]{Batten1978} Batten, A.~H., \& Fletcher, J.~M.\ 1978, \pasp, 90, 312

\bibitem[{{Batten} {et~al.}(1979){Batten}, {Morbey}, {Fekel}, \&  {Tomkin}}]{Batten1979}
{Batten}, A.~H., {Morbey}, C.~L., {Fekel}, F.~C., \& {Tomkin}, J. 1979, \pasp, 91, 304

\bibitem[Batten et al.(1989)]{Batten1989} Batten, A.~H., Fletcher, J.~M., \& MacCarthy,
D.~G.\ 1989, Publications of the Dominion Astrophysical Observatory Victoria, 17, 1

\bibitem[Brancewicz \& Dworak(1980)]{Brancewicz1980} Brancewicz, H.~K., \& Dworak, T.~Z.\ 1980, Acta Astronomica, 30, 501

\bibitem[{{Brettman et al.}(1983)}]{1983IBVSBrettman} Brettman, O.~H., Fried, R.~E., Duvall, W.~M.,
Hall, D.~S., Poe, C.~H., \& Shaw, J.~S.\ 1983, Information Bulletin on Variable Stars, 2389, 1

\bibitem[{{Bruton} {et~al.}(1989){Bruton}, {Hall}, {Boyd}, {Genet}, \&
{Lines}}]{Bruton1989} {Bruton}, J.~R., {Hall}, D.~S., {Boyd}, L.~J., {Genet}, R.~M., \& {Lines},
R.~D. 1989, \apss, 155, 27

\bibitem[Budding(1984)]{Budding1984} Budding, E.\ 1984, Bulletin d'Information du Centre de Donnees Stellaires, 27, 91

\bibitem[Budding et al.(2004)]{Budding2004} Budding, E., Erdem, A., {\c C}i{\c c}ek, C., Bulut, I., Soydugan,
F., Soydugan, E., Baki{\c s}, V., \& Demircan, O.\ 2004, \aap, 417, 263

\bibitem[Carrier et al.(2002)]{Carrier2002} Carrier, F., North, P., Udry, S., \& Babel, J.\ 2002, \aap, 394, 151

\bibitem[Celikel \& Eryurt-Ezer(1989)]{Celikel1989} Celikel, R., \& Eryurt-Ezer, D.\ 1989, \apss, 153, 213

\bibitem[Chamberlin \& McNamara(1957)]{Chamberlin1957} Chamberlin, C., \& McNamara, D.~H.\ 1957, \pasp, 69, 462

\bibitem[{{Chambliss}(1992)}]{Chambliss1992} {Chambliss}, C.~R. 1992, \pasp, 104, 663

\bibitem[{{Chochol} {et~al.}(2006){Chochol}, {Pribulla}, {Va{\v n}ko}, {Mayer},
{Wolf}, {Niarchos}, {Gazeas}, {Manimanis}, {Br{\'a}t}, \& {Zejda}}]{V505Sgr2006} {Chochol}, D.,
{Pribulla}, T., {Va{\v n}ko}, M., {Mayer}, P., {Wolf}, M., {Niarchos}, P.G., {Gazeas}, K.D.,
{Manimanis}, V.N., {Br{\'a}t}, L., {Zejda}, M. 2006, \apss, 304, 93

\bibitem[{{Clausen} {et~al.}(1976){Clausen}, {Gyldenkerne}, \&
{Gronbech}}]{Clausen} {Clausen}, J.~V., {Gyldenkerne}, K., \& {Gronbech}, B. 1976, \aap, 46, 205

\bibitem[Cooper \& Hughes(1994)]{Cooper1994} Cooper, H., \& Hughes, D.~W.\ 1994, \nat, 371, 30

\bibitem[{{Couteau}(1981)}]{Couteau1981} {Couteau}, P. 1981, \aap, 43, 79

\bibitem[Cowley \& Fraquelli(1974)]{Cowley1974PASP} Cowley, A., \& Fraquelli, D.\ 1974, \pasp, 86, 70

\bibitem[Cutispoto et al.(1997)]{Cutispoto1997} Cutispoto, G., Kuerster, M., Messina, S., Rodono, M., \&
Tagliaferri, G.\ 1997, \aap, 320, 586

\bibitem[{{Cvetkovi\'c} {et~al.}(2008)}]{Cvetkovic2008}
{Cvetkovi\'c}, Z., {Novakovi\'c}, B., \& {Todorovi\'c}, N. 2008, New Astronomy 13, 125

\bibitem[{{Demers et al.}(2002)}]{2002Demers} Demers, H., Moffat,
A.~F.~J., Marchenko, S.~V., Gayley, K.~G., \& Morel, T.\ 2002, \apj, 577, 409

\bibitem[Djura{\v s}evi{\'c} et al.(2006)]{Djurasevic2006} Djura{\v s}evi{\'c}, G., Dimitrov, D.,
Arbutina, B., Albayrak, B., Selam, S.~O., \& Atanackovi{\'c}-Vukmanovi{\'c}, O.\ 2006, Publications
of the Astronomical Society of Australia, 23, 154

\bibitem[{{Docobo} \& {Andrade}(2005)}]{Docobo2005}
{Docobo}, J.~A. \& {Andrade}, M. 2005, IAU Commission 26, Inf.\ Circ.\ 157

\bibitem[Docobo \& Andrade(2006)]{Docobo2006} Docobo, J.~A., \& Andrade, M.\ 2006, \apj, 652, 681

\bibitem[{{Docobo} \& {Ling}(2007)}]{Docobo2007LOHya} {Docobo}, J.~A. \& {Ling}, J.~F. 2007, \aj, 133, 1209

\bibitem[{{Docobo} \& {Ling}(2008)}]{Docobo2008}
{Docobo}, J.~A. \& {Ling}, J.~F. 2008, IAU Comm.\ 26, Inf.\ Circ.\ 164

\bibitem[Drechsel et al.(1989)]{Drechsel1989} Drechsel, H., Lorenz, R., \& Mayer, P.\ 1989, \aap, 221, 49

\bibitem[Duerbeck \& Rucinski(2007)]{Duerbeck2007} Duerbeck, H.~W., \& Rucinski, S.~M.\ 2007, \aj, 133, 169

\bibitem[Eaton \& Henry(2007)]{Eaton2007} Eaton, J.~A., \& Henry, G.~W.\ 2007, \pasp, 119, 259

\bibitem[Edwards(1976)]{Edwards1976} Edwards, T.~W.\ 1976, \aj, 81, 245

\bibitem[Erdem et al.(2007)]{Erdem2007NewA} Erdem, A., Soydugan, F.,
Do{\u g}ru, S.~S., {\"O}zkarde{\c s}, B., Do{\u g}ru, D., T{\"u}ys{\"u}z, M., \& Demircan, O.\
2007, New Astronomy, 12, 613

\bibitem[Etzel \& Olson(1985)]{Etzel1985} Etzel, P.~B., \& Olson, E.~C.\ 1985, \aj, 90, 504

\bibitem[Fehrenbach(1961)]{Fehrenbach1961} Fehrenbach, C.\ 1961, Journal des Observateurs, 44, 233

\bibitem[{{Fekel} {et~al.}(1994){Fekel}, {Henry}, {Hampton}, {Fried}, \&
{Morton}}]{Fekel1994} {Fekel}, F.~C., {Henry}, G.~W., {Hampton}, M.~L., {Fried}, R., \& {Morton},
M.~D. 1994, \aj, 108, 694

\bibitem[Ferrer \& Sahade(1986)]{Ferrer1986} Ferrer, O.~E., \& Sahade, J.\ 1986, \pasp, 98, 1342

\bibitem[{{Finsen}(1964)}]{Finsen1964V831Cen}
{Finsen}, W. 1964, Republic Observatory Johannesburg Circulars, 7

\bibitem[Frieboes(1962)]{Frieboes1962} Frieboes, H.~O.\ 1962, \apj, 135, 762

\bibitem[{{Garcia \& Gimenez}(1986)}]{Garcia1986} Garcia, J.~M., \& Gimenez, A.\ 1986, \apss, 125, 181

\bibitem[{{Gim\'{e}nez} {et~al.}(1986){Gim\'{e}nez}, {Clausen}, \& {Jensen}}]{Gimen1986}
{Gim\'{e}nez}, A., {Clausen}, J.~V., \& {Jensen}, K.~S. 1986, \aap, 159, 157

\bibitem[Gimenez \& Clausen(1994)]{Gimenez1994} Gimenez, A., \& Clausen, J.~V.\ 1994, \aap, 291, 795

\bibitem[Ginestet \& Carquillat(2002)]{Ginestet2002} Ginestet, N., \& Carquillat, J.~M.\ 2002, \apjs, 143, 513

\bibitem[Grenier et al.(1999)]{Grenier1999} {Grenier}, S., {Baylac}, M.O., {Rolland}, L., {Burnage},
R., {Arenou}, F., {Briot}, D., {Delmas}, F., {Duflot}, M., {Genty}, V., {G\'omez}, A.E.,
{Halbwachs}, J.L., {Marouard}, M., {Oblak}, E., {Sellier}, A.\ 1999, \aaps, 137, 451

\bibitem[Gray et al.(2006)]{Gray2006AJ} Gray, R.~O., Corbally, C.~J., Garrison, R.~F.,
McFadden, M.~T., Bubar, E.~J., McGahee, C.~E., O'Donoghue, A.~A., \& Knox, E.~R.\ 2006, \aj, 132,
161

\bibitem[{{Griffin}(1999)}]{Griffin1999} {Griffin}, R.~F. 1999, The Observatory, 119, 27

\bibitem[Grosheva(2006)]{Grosheva2006} Grosheva, E.~A.\ 2006, Astrophysics, 49, 397

\bibitem[Guetter(1968)]{Guetter1968} Guetter, H.~H.\ 1968, \pasp, 80, 197

\bibitem[Halbedel(1985)]{Halbedel1985} Halbedel, E.~M.\ 1985, \pasp, 97, 434

\bibitem[Harmanec \& Scholz(1993)]{Harmanec1993} Harmanec, P., \& Scholz, G.\ 1993, \aap, 279, 131

\bibitem[{Hartkopf} {et~al.}(1996){Hartkopf}, {Mason}, \& {McAlister}]{Hartkopf1996} {Hartkopf},
W.~I., {Mason}, B.~D., \& {McAlister}, H.~A. 1996, \aj, 111, 1

\bibitem[{{Harvin et al.}(2002)}]{2002DelOri} Harvin, J.~A., Gies, D.~R., Bagnuolo, W.~G., Jr.,
Penny, L.~R., \& Thaller, M.~L.\ 2002, \apj, 565, 1216

\bibitem[{{Heintz}(1982)}]{Heintz1982} {Heintz}, W.~D., 1982, \pasp, 94, 705

\bibitem[{{Heintz}(1975)}]{Heintz1975} {Heintz}, W.~D., 1975, \apjs, 29, 315

\bibitem[Helt(1987)]{Helt1987} Helt, B.~E.\ 1987, \aap, 172, 155

\bibitem[{{Herczeg} \& {Schmidt}(1960)}]{HerSch}
{Herczeg}, T. \& {Schmidt}, H. 1960, Veroeffentlichungen des Astronomisches
  Institute der Universitaet Bonn, 57, 1

\bibitem[Hilditch(2005)]{Hilditch2005} Hilditch, R.~W.\ 2005, The Observatory, 125, 72

\bibitem[Hill et al.(1975)]{Hill1975} Hill, G., Hilditch, R.~W.,
Younger, F., \& Fisher, W.~A.\ 1975, \memras, 79, 131

\bibitem[{{Hill et al.}(1988)}]{1988Hill} Hill, G.~M., Walker,
G.~A.~H., Dinshaw, N., Yang, S., \& Harmance, P.\ 1988, \pasp, 100, 243

\bibitem[Hinderer(1957)]{Hinderer1957} Hinderer, F.\ 1957, Astronomische Nachrichten, 284, 1

\bibitem[Hoffleit \& Jaschek(1982)]{Hoffleit1982} Hoffleit, D., \& Jaschek, C.\ 1982, The Bright Star
Catalogue, New Haven: Yale University Observatory (4th edition), 1982,

\bibitem[Holmgren(1987)]{Holmgren1987} Holmgren, D.\ 1987, \baas, 19, 709

\bibitem[Houk \& Cowley(1975)]{Houk1975} Houk, N., \& Cowley, A.~P.\ 1975,
Ann Arbor: University of Michigan, Departement of Astronomy, 1975,

\bibitem[Houk(1978)]{Houk1978} Houk, N.\ 1978, Ann Arbor : Dept.~of Astronomy,
University of Michigan : distributed by University Microfilms International, 1978

\bibitem[Houk(1982)]{Houk1982} Houk, N.\ 1982, Michigan Spectral
Survey, Ann Arbor, Dep.~Astron., Univ.~Michigan, 3 (1982), 0

\bibitem[Houk \& Smith-Moore(1988)]{Houk1988} Houk, N., \& Smith-Moore, M.\ 1988, Michigan Spectral
Survey, Ann Arbor, Dept.~of Astronomy, Univ.~Michigan (Vol.~4) (1988), 0

\bibitem[{{{\.I}bano{\v g}lu} {et~al.}(2000){{\.I}bano{\v g}lu}, {{\c C}akirli}, {De{\v g}irmenci},
{Saygan}, {Ula{\c s}}, \& {Erkan}}]{Iban2000} {{\.I}bano{\v g}lu}, C., {{\c C}akirli}, {\"O}.,
{De{\v g}irmenci}, {\"O}., {Saygan}, S., {Ula{\c s}}, B., {Erkan}, N. 2000, \aap, 354, 188

\bibitem[{{Irwin}(1959)}]{Irwin1959} {Irwin}, J.~B. 1959, \aj, 64, 149

\bibitem[Kaszas et al.(1998)]{Kaszas1998} Kaszas, G., Vinko, J., Szatmary, K.,
Hegedus, T., Gal, J., Kiss L.~L., \& Borkovits, T.\ 1998, \aap, 331, 231

\bibitem[{{Kellerer} {et~al.}(2007){Kellerer}, {Petr-Gotzens}, {Kervella}, \&
  {Coud{\'e} Du Foresto}}]{Kellerer2007DelVel}
{Kellerer}, A., {Petr-Gotzens}, M.~G., {Kervella}, P., \& {Coud{\'e} Du
  Foresto}, V. 2007, \aap, 469, 633

\bibitem[Kopal(1978)]{Kopal1978} Kopal, Z.\ 1978, Dordrecht,
D.~Reidel Publishing Co.~(Astrophysics and Space Science Library.~Volume 68), 1978.~524 p.,

\bibitem[K{\H o}v{\'a}ri et al.(2007)]{Kovari2007} K{\H o}v{\'a}ri, Z., Bartus, J.,
Strassmeier, K.~G., Ol{\'a}h, K., Weber, M., Rice, J.~B., \& Washuettl, A.\ 2007, \aap, 463, 1071

\bibitem[{{Kreiner et al.}(2001)}]{Kreiner2001} Kreiner, J.~M., Kim, C.-H.,
 \& Nha, I.-S.\ 2001, An Atlas of O-C Diagrams of Eclipsing Binary Stars / by
 Jerzy M.~Kreiner, Chun-Hwey Kim, Il-Seong Nha.~Cracow, Poland: Wydawnictwo
 Naukowe Akademii Pedagogicznej.~2001.

\bibitem[Krylov et al.(2003)]{Krylov2003} Krylov, A.~V., Mossakovskaya, L.~V.,
Khaliullin, K.~F., \& Khaliullina, A.~I.\ 2003, Astronomy Reports, 47, 551

\bibitem[{{Kwee \& van Woerden}(1956)}]{Kwee} Kwee, K.~K., \& van Woerden, H.\ 1956, \bain, 12, 327

\bibitem[{{Labeyrie} {et al.}(1974)}]{Labeyrie1974} Labeyrie, A., Bonneau, D., Stachnik, R.~V.,
\& Gezari, D.~Y.\ 1974, \apj, 194, L147

\bibitem[Lacy \& Frueh(1985)]{Lacy1985} Lacy, C.~H., \& Frueh, M.~L.\ 1985, \apj, 295, 569

\bibitem[Lacy et al.(1999)]{Lacy1999} Lacy, C.~H.~S., Helt, B.~E., \& Vaz, L.~P.~R.\ 1999, \aj, 117, 541

\bibitem[Lestrade et al.(1993)]{Lestrade1993} Lestrade, J.-F.,
Phillips, R.~B., Hodges, M.~W., \& Preston, R.~A.\ 1993, \apj, 410, 808

\bibitem[Leung \& Schneider(1978)]{Leung1978} Leung, K.-C., \& Schneider, D.~P.\ 1978, \aj, 83, 618

\bibitem[Leung et al.(1979)]{Leung1979} Leung, K.-C., Moffat, A.~F.~J., \& Seggewiss, W.\ 1979, \apj, 231, 742

\bibitem[Levato(1975)]{Levato1975} Levato, H.\ 1975, \aaps, 19, 91

\bibitem[Lindroos(1985)]{Lindroos1985} Lindroos, K.~P.\ 1985, \aaps, 60, 183

\bibitem[Lindroos(1986)]{Lindroos1986} Lindroos, K.~P.\ 1986, \aap, 156, 223

\bibitem[Lippky \& Marx(1994)]{Lippky1994} Lippky, B., \& Marx, S.\ 1994,
Astronomische Gesellschaft Abstract Series, 10, 151

\bibitem[Lorenz et al.(1998)]{Lorenz1998} Lorenz, R., Mayer, P., \& Drechsel, H.\ 1998, \aap, 332, 909

\bibitem[Lu \& Rucinski(1993)]{Lu1993} Lu, W.-X., \& Rucinski, S.~M.\ 1993, \aj, 106, 361

\bibitem[Lu et al.(2001)]{Lu2001} Lu, W., Rucinski, S.~M., \& Og{\l}oza, W.\ 2001, \aj, 122, 402

\bibitem[Malaroda(1975)]{Malaroda1975} Malaroda, S.\ 1975, \aj, 80, 637

\bibitem[Malkov et al.(2006)]{Malkov2006} Malkov, O.~Y., Oblak, E., Snegireva,
E.~A., \& Torra, J.\ 2006, \aap, 446, 785

\bibitem[{{Mason \& Hartkopf}(1999)}]{1999Mason} Mason, B.~D., \& Hartkopf, W.~I.\ 1999, IAU Comm.\ 26, Inf.\ Circ.\ 138

\bibitem[{{Mason} {et~al.}(2001){Mason}, {Wycoff}, {Hartkopf}, {Douglass}, \&
{Worley}}]{WDS} {Mason}, B.~D., {Wycoff}, G.~L., {Hartkopf}, W.~I., {Douglass}, G.~G., \& {Worley},
C.~E. 2001, \aj, 122, 3466

\bibitem[{{Massarotti et al.}(2008)}]{VVCrv2008} Massarotti, A.,
Latham, D.~W., Stefanik, R.~P., \& Fogel, J.\ 2008, \aj, 135, 209

\bibitem[{{Mayer}(1990)}]{Mayer1990}
{Mayer}, P. 1990, Bulletin of the Astronomical Institutes of Czechoslovakia,
  41, 231

\bibitem[{{Mayer}(1997)}]{Mayer1997}
{Mayer}, P. 1997, \aap, 324, 988

\bibitem[{{Mayer}(2004)}]{Mayer2004}
{Mayer}, P. 2004, in ASP Conf. Ser. 318: Spectroscopically and Spatially
  Resolving the Components of the Close Binary Stars, ed. R.~W. {Hilditch},
  H.~{Hensberge}, \& K.~{Pavlovski}, 233--241

\bibitem[{{Mayer} {et~al.}(1992){Mayer}, {Lorenz}, \&
{Drechsel}}]{Mayer1992V871Cen} {Mayer}, P., {Lorenz}, R., \& {Drechsel}, H. 1992, Informational
Bulletin on Variable Stars, 3765, 1

\bibitem[{{McAlister} {et~al.}(1983){McAlister}, {Hendry}, {Hartkopf}, {Campbell}, \& {Fekel}}]{McAlister1983}
{McAlister}, H.~A., {Hendry}, E.~M., {Hartkopf}, W.~I., {Campbell}, B.~G., \& {Fekel}, F.~C. 1983, \apjs, 51, 309

\bibitem[Morgan et al.(1955)]{Morgan1955} Morgan, W.~W., Code,
A.~D., \& Whitford, A.~E.\ 1955, \apjs, 2, 41

\bibitem[{{Murdoch} {et~al.}(1994){Murdoch}, {Hearnshaw}, {Kilmartin}, \&
{Gilmore}}]{Murdoch1994GTMus} {Murdoch}, K.~A., {Hearnshaw}, J.~B., {Kilmartin}, P.~M., \&
{Gilmore}, A.~C. 1994, Experimental Astronomy, 5, 195

\bibitem[Murdoch et al.(1995)]{Murdoch1995} Murdoch, K.~A.,
Hearnshaw, J.~B., Kilmartin, P.~M., \& Gilmore, A.~C.\ 1995, \mnras, 276, 836

\bibitem[{{Muterspaugh} {et~al.}(2006){Muterspaugh}, {Lane}, {Konacki},
{Burke}, {Colavita}, {Kulkarni}, \& {Shao}}]{Muterspaugh2006} {Muterspaugh}, M.~W., {Lane}, B.~F.,
{Konacki}, M., {Burke}, B.F., {Colavita}, M.M., {Kulkarni}, S.R., {Shao}, M. 2006, \aap, 446, 723

\bibitem[{{Muterspaugh} {et~al.}(2008){Muterspaugh}, {Lane}, {Fekel},
{Konacki}, {Burke}, {Kulkarni}, {Colavita}, {Shao}, \& {Wiktorowicz}}]{Muterspaugh2008}
{Muterspaugh}, M.W., {Lane}, B.F., {Fekel}, F.C., {Konacki}, M., {Burke}, B.F., {Kulkarni}, S.R.,
{Colavita}, M.M., {Shao}, M., {Wiktorowicz}, S.J.\ 2008, \aj, 135, 766

\bibitem[{{Newburg}(1969)}]{1969Newburg} Newburg, J.~L.\ 1969, Republic
Observatory Johannesburg Circular, 7, 190

\bibitem[{{Olevi{\'c}}(2002)}]{Olevic2002V348And}
{Olevi{\'c}}, D. 2002, IAU Comm.\ 26, Inf.\ Circ.\ 147

\bibitem[{{Olsen}(1972)}]{1972Olsen} Olsen, E.~H.\ 1972, \aap, 20, 167

\bibitem[Olson(1982)]{Olson1982} Olson, E.~C.\ 1982, \apj, 257, 198

\bibitem[{{Otero}(2007)}]{2007OEJV72}
{Otero}, S.~A. 2007, Open European Journal on Variable Stars, 72, 1

\bibitem[{{Otero} {et~al.}(2000){Otero}, {Fieseler}, \& {Lloyd}}]{Otero2000} {Otero}, S.~A.,
{Fieseler}, P.~D., \& {Lloyd}, C. 2000, Informational Bulletin on Variable Stars, 4999, 1

\bibitem[Pagel(1960)]{Pagel1960} Pagel, B.~E.~J.\ 1960, \aj, 65, 352

\bibitem[{{Pan} {et~al.}(1993){Pan}, {Shao}, \& {Colavita}}]{Pan1993Algol}
{Pan}, X., {Shao}, M., \& {Colavita}, M.~M. 1993, \apjl, 413, L129

\bibitem[{{Parsons}(2004)}]{Parsons2004} {Parsons}, S.~B. 2004, \aj, 127, 2915

\bibitem[Pavlovski et al.(1992)]{Pavlovski1992} Pavlovski, K., Schneider, H., \& Akan, M.~C.\ 1992, \aap, 258, 329

\bibitem[{{Pavlovski et al.}(1997)}]{1997Pavlovski} Pavlovski, K., Harmanec, P., Bozic, H., Koubsky, P., Hadrava, P.,
Kriiz, S., Ruzic, Z., \& Stefl, S.\ 1997, \aaps, 125, 75

\bibitem[{{Perryman} \& {ESA}(1997)}]{HIP}
{Perryman}, M.~A.~C. \& {ESA}. 1997, {The HIPPARCOS and TYCHO catalogues} (The
  Hipparcos and Tycho catalogues.~Astrometric and photometric star catalogues
  derived from the ESA Hipparcos Space Astrometry Mission, Publisher:
  Noordwijk, Netherlands: ESA Publications Division, 1997, Series: ESA SP
  Series 1200)

\bibitem[Petrie(1950)]{Petrie1950} Petrie, R.~M.\ 1950,
Publications of the Dominion Astrophysical Observatory Victoria, 8, 319

\bibitem[Plavec et al.(1982)]{Plavec1982} Plavec, M.~J., Weiland, J.~L., \& Koch, R.~H.\ 1982, \apj, 256, 206

\bibitem[Plavec(1983)]{Plavec1983} Plavec, M.~J.\ 1983, \apj, 275, 251

\bibitem[Polidan \& Plavec(1984)]{Polidan1984} Polidan, R.~S., \& Plavec, M.~J.\ 1984, \aj, 89, 1721

\bibitem[{{Popovi{\'c}} \& {Pavlovi\'c}(1995)}]{Popovic1995V773Cas}
{Popovi{\'c}}, G.~M. \& {Pavlovi\'c}, R. 1995, Bulletin Astronomique de Belgrade, 151, 45

\bibitem[Popper(1957)]{Popper1957} Popper, D.~M.\ 1957, \apj, 126, 53

\bibitem[Popper(1981)]{Popper1981} Popper, D.~M.\ 1981, \pasp, 93, 318

\bibitem[{{Popper}(1986)}]{Popper1986} Popper, D.~M.\ 1986, \pasp, 98, 1312

\bibitem[{{Pribulla} {et al.}(2006)}]{Pribulla2006} {Pribulla}, T., {Rucinski}, S.M., {Lu}, W.,
{Mochnacki}, S.W., {Conidis}, G., {Blake}, R.M., {DeBond}, H., {Thomson}, J.R., {Pych}, W.,
{Og{\l}oza}, W., {Siwak}, M.\ 2006, \aj, 132, 769

\bibitem[Proust et al.(1981)]{Proust1981} Proust, D., Ochsenbein, F., \& Pettersen, B.~R.\ 1981, \aaps, 44, 179

\bibitem[Pych et al.(2004)]{Pych2004} {Pych}, W., {Rucinski}, S.M., {DeBond}, H., {Thomson}, J.R.,
{Capobianco}, C.C., {Blake}, R.M., {Og{\l}oza}, W., {Stachowski}, G., {Rogoziecki}, P., {Ligeza},
P., {Gazeas}, K.\ 2004, \aj, 127, 1712

\bibitem[Reglero et al.(1991)]{Reglero1991} Reglero, V., Fernandez-Figuerora, M.~J., Gimenez,
A., de Castro, E., Fabregat, J., Cornide, M., \& Armentia, J.~E.\ 1991, \aaps, 88, 545

\bibitem[Ribas et al.(1999)]{Ribas1999} Ribas, I., Jordi, C., \& Torra, J.\ 1999, \mnras, 309, 199

\bibitem[Roberts et al.(2005)]{Roberts2005} {Roberts}, Jr., L.C., {Turner}, N.H., {Bradford}, L.W.,
{ten Brummelaar}, T.A., {Oppenheimer}, B.R., {Kuhn}, J.R., {Whitman}, K., {Perrin}, M.D., {Graham},
J.R.\ 2005, \aj, 130, 2262

\bibitem[Rucinski et al.(1993)]{Rucinski1993} Rucinski, S.~M., Lu, W.-X., \& Shi, J.\ 1993, \aj, 106, 1174

\bibitem[{{Rucinski} {et~al.}(2002){Rucinski}, {Lu}, {Capobianco}, {Mochnacki}, {Blake},
{Thomson}, {Og{\l}oza}, \& {Stachowski}}]{Rucinski2002} {Rucinski}, S.M., {Lu}, W.,
{Capobianco}, C.C., {Mochnacki}, S.W., {Blake}, R.M., {Thomson}, J.R., {Og{\l}oza}, W.,
{Stachowski}, G.\ 2002, \aj, 124, 1738

\bibitem[Rucinski et al.(2007)]{Rucinski2007} Rucinski, S.~M.,
Pribulla, T., \& van Kerkwijk, M.~H.\ 2007, \aj, 134, 2353

\bibitem[Rucinski et al.(2008)]{Rucinski2008} {Rucinski}, S.M., {Pribulla}, T., {Mochnacki}, S.W.,
{Liokumovich}, E., {Lu}, W., {DeBond}, H., {de Ridder}, A., {Karmo}, T., {Rock}, M., {Thomson},
J.R., {Og{\l}oza}, W., {Kaminski}, K., {Ligeza}, P.\ 2008, \aj, 136, 586

\bibitem[Samec et al.(1989)]{Samec1989} Samec, R.~G., Fuller,
R.~E., \& Kaitchuck, R.~H.\ 1989, \aj, 97, 1159

\bibitem[Samus et al.(2004)]{GCVS2004} Samus, N.~N., Durlevich,
O.~V., \& et al.\ 2004, VizieR Online Data Catalog, 2250, 0

\bibitem[Sarma \& Radharkrishnan(1982)]{Sarma1982} Sarma, M.~B.~K., \& Radharkrishnan, K.~R.\ 1982,
Information Bulletin on Variable Stars, 2073, 1

\bibitem[Schr{\"o}der \& Schmitt(2007)]{Sch2007} Schr{\"o}der, C., \& Schmitt, J.~H.~M.~M.\ 2007, \aap, 475,
677

\bibitem[{{Seymour}(2001)}]{Seymour2001ETBoo}
{Seymour}, D. 2001, IAU Comm.\ 26, Inf.\ Circ.\ 145

\bibitem[{{Seymour} {et~al.}(2002){Seymour}, {Mason}, {Hartkopf}, \&
{Wycoff}}]{Seymour2002V2083Cyg} {Seymour}, D.~M., {Mason}, B.~D., {Hartkopf}, W.~I., \& {Wycoff},
G.~L. 2002, \aj, 123, 1023

\bibitem[Shobbrook(2005)]{Shobbrook2005JAD} Shobbrook, R.~R.\ 2005,
Journal of Astronomical Data, 11, 7

\bibitem[Simonson(1968)]{Simonson1968} Simonson, S.~C.~I.\ 1968, \apj, 154, 923

\bibitem[{{S{\"o}derhjelm}(1999)}]{Soderhjelm1999V640Cas}
{S{\"o}derhjelm}, S. 1999, \aap, 341, 121

\bibitem[{{Stebbins} (1910)}]{Stebbins1910} Stebbins, J.\ 1910, \apj, 32, 185

\bibitem[Stephenson(1960)]{Stephenson1960} Stephenson, C.~B.\ 1960, \aj, 65, 60

\bibitem[{{Sterken}(2005)}]{Sterken2005} Sterken, C.\ 2005, The Light-Time Effect
 in Astrophysics: Causes and cures of the O-C diagram, 335

\bibitem[Stickland et al.(1998)]{Stickland1998} Stickland, D.~J.,
Lloyd, C., \& Sweet, I.\ 1998, The Observatory, 118, 7

\bibitem[Strohmeier(1959)]{Strohmeier1959AN} Strohmeier, W.\ 1959,
Astronomische Nachrichten, 285, 87

\bibitem[Svechnikov \& Kuznetsova(1990)]{Svechnikov1990} Svechnikov, M.~A.,
\& Kuznetsova, E.~F.\ 1990, Sverdlovsk : Izd-vo Ural'skogo universiteta, 1990

\bibitem[{{Tikkanen}(2002)}]{BBSAG125} Tikkanen, K. 2002, BBSAG, 125, 1

\bibitem[{{Tokovinin} {et~al.}(2006){Tokovinin}, {Thomas}, {Sterzik}, \&
{Udry}}]{Tokovinin2006DILyn} {Tokovinin}, A.~A., {Thomas}, S., {Sterzik}, M., \& {Udry}, S. 2006,
\aap, 450, 681

\bibitem[Tokovinin(1997)]{Tokovinin1997MSC} Tokovinin, A.~A.\ 1997, \aaps, 124, 75

\bibitem[Tokovinin(1998)]{Tokovinin1998AstL} Tokovinin, A.~A.\ 1998,
Astronomy Letters, 24, 288

\bibitem[{{Tokovinin} {et~al.}(1999)}]{Tokovinin1999}
{Tokovinin}, A.~A., Chalabaev, A., Shatsky, N.~I., \& Beuzit, J.~L. 1999, \aap, 346, 481

\bibitem[{{Tomkin}(1992)}]{Tomkin1992} {Tomkin}, J. 1992, \apj, 387, 631

\bibitem[Torres(2001)]{Torres2001} Torres, G.\ 2001, \aj, 121, 2227

\bibitem[van Hamme et al.(1994)]{vanHamme1994} {van Hamme}, W.~V., {Hall}, D.~S., {Hargrove},
A.~W., {Henry}, G.~W., {Wasson}, R., {Barkslade}, W.~S., {Chang}, S., {Fried}, R.~E., {Green},
C.~L., {Lines}, H.~C., {Lines}, R.~D., {Nielsen}, P., {Powell}, H.~D., {Reisenweber}, R.~C.,
{Rogers}, C.~W., {Shervais}, S., {Tatum}, R.\ 1994, \aj, 107, 1521

\bibitem[{{van Leeuwen} \& {van Genderen}(1997)}]{Leeuwen1997TauCMa}
{van Leeuwen}, F. \& {van Genderen}, A.~M. 1997, \aap, 327, 1070

\bibitem[Vaz \& Andersen(1984)]{Vaz1984} Vaz, L.~P.~R., \& Andersen, J.\ 1984, \aap, 132, 219

\bibitem[Walborn(1973a)]{Walborn1973} Walborn, N.~R.\ 1973a, \aj, 78, 1067

\bibitem[Walborn(1973b)]{Walborn1973II} Walborn, N.~R.\ 1973b, \apj, 179, 517

\bibitem[Walker \& Chambliss(1985)]{Walker1985} Walker, R.~L., \& Chambliss, C.~R.\ 1985, \aj, 90, 346

\bibitem[{{Watson et al.}(2001)}]{2001BBScl} Watson, L.~C., Pritchard, J.~D., Hearnshaw, J.~B.,
Kilmartin, P.~M., \& Gilmore, A.~C.\ 2001, \mnras, 325, 143

\bibitem[{{Wolf} \& {Caffey}(1998)}]{ObjevDILyn1998}
{Wolf}, G.~W. \& {Caffey}, J.~F. 1998, Informational Bulletin on Variable
  Stars, 4649, 1

\bibitem[{{Wolf}(2000)}]{Wolf2000V1647Sgr}
{Wolf}, M. 2000, \aap, 356, 134

\bibitem[{{Wolf} {et~al.}(2006){Wolf}, {Ku{\v c}{\'a}kov{\'a}}, {Kolasa}, {{\v S}tastn{\'y}},
{Bozkurt}, {Harmanec}, {Zejda}, {Br{\'a}t}, \& {Hornoch}}]{Wolf2006AGPer} {Wolf}, M., {Ku{\v
c}{\'a}kov{\'a}}, H., {Kolasa}, M., {{\v S}tastn{\'y}}, P., {Bozkurt}, Z., {Harmanec}, P., {Zejda},
M., {Br{\'a}t}, L., {Hornoch}, K.\ 2006, \aap, 456, 1077

\bibitem[Wolf et al.(2008)]{Wolf2008} Wolf, M., Zejda, M.,
\& de Villiers, S.~N.\ 2008, \mnras, 388, 1836

\bibitem[{{Woodward} \& {Koch}(1987)}]{Woodward1987}
{Woodward}, E.~J. \& {Koch}, R.~H. 1987, \apss, 129, 187

\bibitem[{{Yakut} {et~al.}(2004){Yakut}, {Kalomeni}, \& {{\.I}bano{\u g}lu}}]{Yakut2004}
{Yakut}, K., {Kalomeni}, B., \& {{\.I}bano{\u g}lu}, C. 2004, \aap, 417, 725

\bibitem[{{Zaera}(1985)}]{Zaera85} {Zaera}, J.~A. 1985, IAU Comm.\ 26, Inf.\ Circ.\ 96

\bibitem[{{Zasche}(2008)}]{Zasche2008} Zasche, P.\ 2008, arXiv:0801.4258

\bibitem[{{Zasche} \& {Svoboda}(2008)}]{IBVSV348And}
{Zasche}, P. \& {Svoboda}, P. 2008, Informational Bulletin on Variable Stars,
  in prep.

\bibitem[{{Zasche} \& {Wolf}(2007)}]{ZascheWolf}
{Zasche}, P. \& {Wolf}, M. 2007, Astronomische Nachrichten, 328, 928

\bibitem[Zejda et al.(2008)]{Zejda2008} Zejda, M., Mikul{\'a}{\v s}ek, Z., \& Wolf, M.\ 2008, \aap, 489, 321

\bibitem[Zhang \& Gu(2008)]{Zhang2008} Zhang, L.-Y., \& Gu, S.-H.\ 2008, \aap, 487, 709

\end{thebibliography}
\end{document}